\documentclass[10pt,a4paper,floatfix]{article}
\usepackage{naturetex}
\usepackage[singlelinecheck=false]{caption}
\usepackage{booktabs}
\usepackage{bm}
\usepackage{multicol}
\usepackage{comment}
\usepackage[superscript,biblabel]{cite}
\usepackage{amsmath,amssymb}
\usepackage{tablefootnote}
\usepackage{array}
\newcolumntype{H}{>{\setbox0=\hbox\bgroup}c<{\egroup}@{}}
 \usepackage{hyperref}
 \hypersetup{
    bookmarks=true,         % show bookmarks bar?
    unicode=false,          % non-Latin characters in AcrobatÕs bookmarks
    pdftoolbar=true,        % show AcrobatÕs toolbar?
    pdfmenubar=true,        % show AcrobatÕs menu?
    pdffitwindow=false,     % window fit to page when opened
    pdfstartview={FitH},    % fits the width of the page to the window
    pdftitle={Many-body theory of positron binding to polyatomic molecules},   % title
    pdfauthor={D. G. Green},     % author
    pdfsubject={PsAtomMBT},   % subject of the document
    pdfcreator={D. G. Green},   % creator of the document
    pdfproducer={Producer}, % producer of the document
    pdfkeywords={keywords}, % list of keywords
    pdfnewwindow=true,      % links in new window
    colorlinks=true,       % false: boxed links; true: colored links
    linkcolor=blue,          % color of internal links
    citecolor=blue,        % color of links to bibliography
    filecolor=magenta,      % color of file links
    urlcolor=blue           % color of external links
}

\newcommand{\red}[1]{{\textcolor{red}{#1}}}%%
\newcommand{\blue}[1]{{\textcolor{blue}{#1}}}%%
\renewcommand{\figurename}{\footnotesize{Fig.}}

\begin{document}
\maketitle

\setlength{\abovedisplayskip}{5.3pt}
\setlength{\belowdisplayskip}{5.3pt}

\noindent{{\bf 
\noindent Positrons bind to molecules leading to vibrational excitation and spectacularly enhanced annihilation\cite{RevModPhys.82.2557}. 
Whilst positron binding energies have been measured via resonant annihilation spectra for $\sim$ 90 molecules in the past two decades\cite{Gilbert02,Danielson09,Danielson10,Danielson12,Danielson12a}, an accurate \emph{ab initio} theoretical description has remained elusive. Of the molecules studied experimentally, calculations exist for only 6, and 
for these, standard quantum chemistry approaches have proved severely deficient, agreeing with experiment to at best 25\% accuracy for polar molecules, and failing to predict binding in nonpolar molecules. %\cite{Surko88,Gilbert02,Barnes03,Barnes06,Young07,Young08_large,Young08_small,}
%The failure to describe such a fundamental system is unsatisfactory and surprising, especially given the maturity and success of electronic structure theory. 
The mechanisms of binding are not understood. 
%The theoretical difficulty lies in the need to system.
Here, we develop a many-body theory of positron-molecule interactions 
%and elucidate the mechanisms of positron binding, 
and uncover the role of strong many-body
%identify and accurately account for the strong many-body correlations that characterize the positron-molecule 
%We identify  %takes \emph{ab initio} account of %and elucidates the role of
%The theoretical difficulty lies in the need to accurately account for 
%positron-molecule correlations including 
correlations including polarization of the electron cloud, screening of the positron-molecule Coulomb interaction by molecular electrons, and crucially, the  unique non-perturbative process of virtual-positronium formation (where a molecular electron temporarily tunnels to the positron): they dramatically enhance binding in polar molecules and enable binding in nonpolars. % it dramatically enhances binding in polar molecules, and can be essential to enable binding in nonpolar molecules. 
%Their roles in positron-molecule binding have yet to be elucidated. 
%%and delineates the effects 
%of the correlations.
% for the molecules for which both theory and experimental binding energies exist, including non-polars for which binding is enabled exclusively by correlations.
%we find agreement with experiment to within 10\% in cases. 
% and we predict binding in formamide.
%We uncover the crucial role of virtual-positronium formation: 
We also elucidate the role of individual molecular orbitals, highlighting the importance of electronic $\pi$ bonds. Overall, we calculate binding energies in agreement with experiment (to within 1\% in cases), and we predict binding in formamide and nucleobases.
As well as supporting resonant annihilation experiments and positron-based molecular spectroscopy, the approach can be extended to positron scattering and annihilation $\gamma$ spectra in molecules and condensed matter, to provide fundamental insight and predictive capability required to properly interpret materials science diagnostics \cite{RMPpossolids2013,hugreview}, 
develop antimatter-based technologies (including positron traps, beams and positron emission tomography\cite{Danielson:2015,Natisin:2016,hugreview}), and understand positrons in the galaxy\cite{RevModPhys.83.1001}. %Moreover, the positron-matter problem provides an unforgiving testbed for the development of computational methods to tackle the quantum many-body problem, for which our results can serve as benchmarks.
}}
%\red{Elucidate role of individual MOs to binding, highlighting importance of $\Pi$ bonds, which may contribute to binding more than HOMO\dots}

%\red{Charles' comments:
%\begin{itemize}
%\item 1) Do the quantum chemistry approaches not work because they are limited in excitation level and therefore only get the first couple of virtual positronium diagrams (or maybe just one)? If you feel sure of this you might point this out in the discussion.
%\item (2) When you quote 2 values for sigmaGW+gamma+lambda is that an estimate range or where do the 2 numbers come from? Needs to be made explicit.
%\item (3) You refer to BSE ionization energies in Extended data Table 1.  Usually IPs would come from sigma. I know that you are solving for sigma by a Dyson equation, but putting BSE in there might be confusing for condensed matter people
%\end{itemize}
%}

\vspace*{1ex}
Pioneering technological developments have enabled the trapping, accumulation and delivery \cite{Danielson:2015,Natisin:2016,hugreview} of positrons for study of their fundamental interactions with atoms and molecules\cite{Surko:2005,RevModPhys.82.2557}, and the formation, exploitation and interrogation of 
%more complicated antimatter, namely 
positronium (Ps) \cite{Brawley:2010,cassidy2018}  and antihydrogen\cite{Baker:2021}. 
The ability of positrons to annihilate with atomic electrons forming characteristic $\gamma$ rays makes them a unique probe over vast length scales, giving them important use in e.g., 
materials science for ultra-sensitive diagnostics of industrially important materials\cite{hugreview, RMPpossolids2013}, medical imaging (positron emission tomography) \cite{PETbook}, and in astrophysics\cite{RevModPhys.83.1001}. 

Proper interpretation of the  %difficult and costly antimatter 
%experiments and 
materials science techniques, and the development of next-generation antimatter-based technologies rely on an accurate understanding of the fundamental interactions of positrons with atoms and molecules. 
%Yet, \red{and despite the maturity of the experiments and technologies,} 
Yet, many 
%important and 
basic aspects
% of positron interactions with matter 
 are poorly understood theoretically. % owing to the complexity of antimatter-matter interactions. 
A striking example is the open fundamental problem of positron binding to molecules. 
%The theoretical and computational  \emph{electronic} structure of molecules is key to chemistry and biology. 
Positrons can readily attach to molecules that bind them via vibrational Feshbach resonances, leading to spectacularly enhanced annihilation spectra that exhibit pronounced resonances downshifted from the vibrational mode energies by the positron-molecule binding energy\cite{RevModPhys.82.2557}. 
Observation of such energy-resolved annihilation spectra  have enabled measurement of positron binding energies (ranging from a few to a few hundred meV) for more than 90 molecules. The majority of these ($\sim$60) are nonpolar or weakly polar species, such as alkanes, aromatics, partially halogenated hydrocarbons, alcohols, formates, and acetates. 
 \emph{Ab initio} calculations have been performed predominantly for strongly polar molecules, see e.g., Refs.~[\citen{Danby:1988,STRASBURGER199649,Chojnacki06,PhysRevA.73.022705, bubin,Mella02,Kita09,APMO2014}] (we note there have been recent model calculations\cite{Swann:2019,MLbind,molbind_dft2}).
% , for which binding is guaranteed at even the static level of theory. 
%(i.e., those with a dipole moment greater than the critical value of 1.625\,D that guarantees binding even at the static level of theory\cite{Crawford67}) 
%using a variety of sophisticated methods including the
%$R$-matrix\cite{Danby:1988},
%configuration interaction (CI)\cite{STRASBURGER199649,Chojnacki06,PhysRevA.73.022705}, explicitly correlated Gaussians\cite{bubin}, NEO (Nuclear Electronic Orbital framework)\cite{}, quantum Monte Carlo (QMC) \cite{Mella02,Kita09}, and `any-particle molecular-orbital' (APMO)\cite{APMO2014} approaches. 
Remarkably, only six species have been studied both  experimentally and with \emph{ab initio} theory, namely carbon disulphide CS$_2$, acetaldehyde C$_2$H$_4$O, propanal C$_2$H$_5$CHO, acetone (CH$_3$)$_2$CO, acetonitrile CH$_3$CN, and propionitrile C$_2$H$_5$CN\cite{RevModPhys.82.2557}. 
For these, the sophisticated quantum chemistry approaches proved severely deficient,  the best agreement being $\gtrsim$25\% (for acetonitrile, theory: $\varepsilon_b=136$ meV\cite{Tachikawa14} 
vs experiment: $\varepsilon_b=180$ meV\cite{Danielson12}), and failing to predict binding in nonpolar CS$_2$ (vs.~experiment: $\varepsilon_b=75$ meV \cite{Danielson10}) (see Table 1). % whereas no calculation has yet predicted binding. 
Moreover, it has been observed that positron-molecule binding energies can be significantly larger than electron-molecule ones (i.e., negative ion states) \cite{Danielson10,Danielson12a}, yet the differences are not quantitatively understood.
%\red{[We've lost the ML reference]} %... and Andrew and japanese model ones}
%This disparity is unsatisfactory, and somewhat shocking given the success of \emph{electronic} structure theory.

\begin{figure*}[t!!]
%\centering\includegraphics[width=0.65\textwidth]{diags_jd2}
%\vspace*{-4ex}
\centering\includegraphics[width=0.73\textwidth]{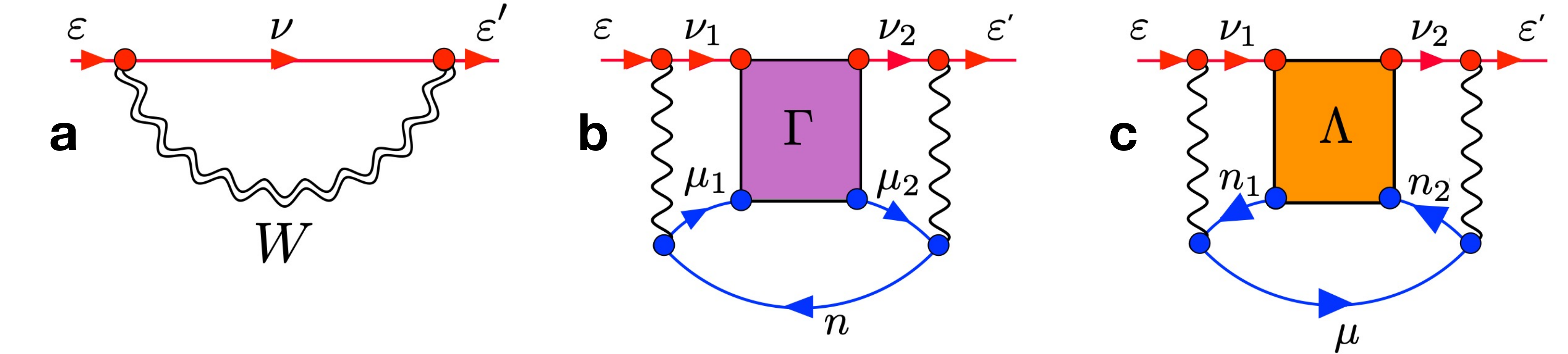}
%\centering\includegraphics[width=0.73\textwidth]{diags-small}
%\centering\includegraphics[width=0.9\textwidth]{diags_jd}
\vspace*{-1ex}
\caption*{ \footnotesize {\bf Fig.~1. Main contributions to the positron-molecule self energy}. 
{\bf a}, the `$GW$' contribution, which involves the positron Green's function $G_{\nu}$ and the (dynamic part of the) screened Coulomb interaction $W$. 
%$W_d=v\Pi v$, where $v$ is a bare Coulomb interaction and $\Pi$ is the electron-hole polarization propagator, which satisfies a Bethe-Salpeter equation (see {\bf b} and {\bf e}). 
%Depending on the approximation of the kernel $K$ of $\Pi$, the 
%$GW$ diagram describes polarization of the molecular electron cloud ($K=0$), or \emph{additionally}, screening of the electron-positron Coulomb interaction by molecular electrons ($K=v$), and electron-hole attractions ($K=v-W$).
%{\bf c}, the virtual-positronium contribution  $\Sigma^{(\Gamma)}$, which includes the summed infinite ladder series  (`$\Gamma$-block', see {\bf g}) of electron-positron interactions.
%{\bf d}, $\Sigma^{(\Lambda)}$, where the `$\Lambda$ block' has similar form to the $\Gamma$-block but is the summed infinite ladder series of positron-hole interactions.
It describes the positron-induced polarization of the molecular electron cloud and corrections to it due to screening of the electron-positron Coulomb interaction by molecular electrons, and electron-hole attractions.
{\bf b}, the virtual-positronium contribution  $\Sigma^{\Gamma}$, which includes the summed infinite ladder series  (`$\Gamma$-block') of screened electron-positron interactions.
{\bf c}, the positron-hole ladder series (the `$\Lambda$ block') contribution $\Sigma^{\Lambda}$. 
%where has similar form to the $\Gamma$-block but is the summed infinite ladder series of positron-hole interactions. 
Lines directed to the right (left) represent particles (holes) propagating on the $N$-electron ground-state molecule: 
% positron ext
red lines labelled $\varepsilon$ represent the external positron state; 
% positron/ectron intermediate
other red (blue) lines represent positron (excited electron or hole) intermediate states that are summed over; 
single (double) wavy lines represent bare (screened) Coulomb interactions. 
%$W$, $\Gamma$ and $\Lambda$ satisfy 
See `Methods' and Extended Data Fig.~1 for details of their calculation via Bethe-Salpeter equations.\\[-4.5ex]}
%{\bf (a)} The $\Sigma^{\rm GW}$ contribution, which represents polarization of the molecular electron cloud by the positron, screening of the electron-positron interaction by molecular electrons and the electron-hole attraction. It is the product of the positron Green's function and the dynamic part (due to the absence of a positron-electron exchange interaction) of the screened electron-positron Coulomb interaction  [double wavy line, diagram (b)], where $\Pi$ is the dressed electron-hole polarization propagator. 
%It satisfies the Bethe-Salpeter equation [diagram (e)] with kernel $K=v-W_{\rm RPA}$ [diagram (f)], where $W_{\rm RPA}= v+ W_{d, \rm RPA}$ is the full screened electron-hole Coulomb interaction calculated in the random phase approximation. %with $W_d$ calculated using the polarization propagator kernel $K=v$, 
%{\bf (c)} The virtual-positronium contribution  $\Sigma^{\Gamma}$, which includes the summed infinite ladder-diagram series of screened electron-positron interactions,  the `${\Gamma}$ block' [diagram (g)] and (d) the positron-hole ladder series contribution. 
%See also Extended Data Fig.~1.}
\end{figure*}

The theoretical difficulty lies in the need to identify and accurately describe the strong many-body correlations that characterise the positron-molecule system. %, whose
%roles in positron-molecule binding have yet to be elucidated.
Here, we develop the \emph{ab initio} many-body theory of positron interactions with polyatomic molecules, which enables the natural, explicit, and systematic account of important correlations including
polarization of the molecular electron cloud, screening of the electron-positron interaction,  
%by molecular electrons, 
and the unique process of virtual-positronium formation (where a molecular electron temporarily tunnels to the positron). %, occurring when the positron energy is lower than the Ps-formation threshold. 
%$\varepsilon_{\rm Ps} = I + E_{\rm 1s}({\rm Ps}) = I - 6.8$ eV, where $I$ is the ionization potential and $E_{\rm 1s}$(Ps) is the ground-state energy of Ps). 
%They make the theoretical description of positron-molecule interactions a challenging many-body problem, and 
% and for molecules, the different length scales over which they act also bring formidable computational challenges.
 %nor delineated. 
% which takes full account of the important correlations, and enables delineation of their effects.
%It takes full account of the important correlations in a natural, intuitive and systematically improvable way, is capable of treating binding, scattering and annihilation on the same footing, and is scalable to molecules with $>10$ atoms.
% in a natural, intuitive and systematically improvable way, is capable of treating binding, scattering and annihilation on the same footing, 
% and is scalable to molecules with $>10$ atoms.
We focus its application, via a state-of-the-art computational implementation, to calculations of positron binding and annihilation
%bound-state lifetimes (against annihilation) 
for the six molecules for which both previous theory and measurements exist, and additionally formamide, CSe$_2$, benzene and A, C, T, G and U nucleobases.
The positron binding energy $\varepsilon$ and bound-state wavefunction $\psi_{\varepsilon}$  are found by solving the Dyson equation\cite{mbtexposed} 
\begin{eqnarray}\label{eqn:dyson}
\left(H^{(0)} + {\Sigma}_{\varepsilon}\right)\psi_{\varepsilon}({\bf r}) = \varepsilon\psi_{\varepsilon}({\bf r}),
\end{eqnarray}
where ${H}^{(0)}$ is the 
%zeroth-order 
Hamiltonian 
%that is taken to be that 
of the positron in the Hartree-Fock (HF) field of the ground-state molecule, and 
 ${\Sigma}_{E}$ is a non-local, energy-dependent correlation potential (irreducible self energy of the positron). 
%  in the field of the molecule) 
%  \cite{PhysRevLett.3.96}, 
It acts as an integral operator $\Sigma_E\psi({\bf r}) \equiv \int \Sigma_E({\bf r,r}')\psi({\bf r}')d{\bf r}'$ and encapsulates the full complexity of the many-body problem.  
We calculate 
%matrix elements of 
$\Sigma$ %in the positron HF molecular-orbital (MO) basis 
via its expansion in residual electron-electron and electron-positron interactions, see Fig.~1. % and Extended Data Fig.~1. % (in practice computing its (see `Methods'). 
Diagram (a), the `$GW$' self energy $\Sigma^{GW}$, 
%involves the positron Green's function $G$ and the screened electron-positron Coulomb interaction $W=v\Pi v$, where $v$ is the bare Coulomb interaction and $\Pi$ is the dressed polarization propagator (see Extended Fig). %involves the positron Green's function $G$ and the screened positron-electron Coulomb interaction $W$. It 
describes the positron-induced polarization of the molecular electron cloud, 
%(denoted $GW@\Sigma^{(2)}$) and can additionally include 
and corrections to it due to screening of the electron-positron Coulomb interaction by the molecular electrons, 
%(the \emph{random phase approximation}, `$GW$@RPA'), 
and electron-hole attractions (the \emph{Bethe-Salpeter Equation approximation}, `$GW$@BSE'). 
Diagram (b), denoted by $\Sigma^{\Gamma}$,
 which involves the summed infinite ladder series of (screened) electron-positron interactions (the `$\Gamma$-block', see Extended Data Fig.~1), % [diagram (f)]. %It satisfies a linear integral equation. 
represents virtual-positronium formation \cite{dzuba_mbt_noblegas,DGG_posnobles}. %,DGG:2015:core
%It is unique to the positron-atom and positron-molecule problem, 
Its importance is unique to the positron problem because successive diagrams in this series contribute to the positron-molecule self energy with the same sign, 
%yielding an attractive interaction whose strength is comparable to (or greater than) the polarization potential, 
whereas for all-electron systems the series is sign alternating and gives a small overall contribution. 
%Its contribution to positron-molecule interactions is unknown prior to this work.
We also consider the smaller positron-hole ladder series contribution, $\Sigma^{\Lambda}$, Fig.~1\,c.
`Methods' details the state-of-the-art construction of $\Sigma$ and solution of the Dyson equation. % are detailed in .%This is similar in structure to diagram (b) but instead involves . %which satisfies a similar linear integral equation to $\Gamma$. 

\section*{Results and discussion: positron binding energies and lifetimes.}
\vspace*{-1ex}
%\subsubsection*{Positron binding to polar molecules}
%\noindent{\bf \emph{Positron binding in polar molecules}}.---
%\noindent{\bf {Positron binding in polar and nonpolar molecules}}.

%\smallskip
%\smallskip
\noindent Table 1 shows our calculated binding energies at successively more sophisticated approximations to the correlation potential: HF; $\Sigma^{(2)}$ (bare polarisation); $\Sigma^{GW}$ (polarisation including electron screening and screened electron-hole interactions (Fig.~1\,a); 
% in its various levels, $\Sigma^{(GW)}+\Sigma^{(\Gamma)}$, 
$\Sigma^{GW+\Gamma}$ (Fig.~1 a+b); 
$\Sigma^{GW+\Gamma+\Lambda}$ (Fig.~1 a+b+c): for this the first (second) number is the result using bare (dressed) Coulomb interactions in the ladders, whilst the third (our most sophisticated, in bold) is that using dressed interactions and energies. Also see Fig.~2 for a graphical comparison of theory and experiment, and Extended Data Table 2 for more details. % and that with screened interactions in the ladders, denoted by tildes). 
% for a number of polar organic and non-polar molecules, including the six molecules for which both theory and experimental results exist, and additionally LiH, formamide, CSe$_2$ and benzene. 
%Our best results(most sophisticated approximation) are the $\Sigma^{GW}+\Sigma^{\Gamma}+\Sigma^{\Lambda}$ and $\Sigma^{GW}+\Sigma^{\tilde{\Gamma}}+\Sigma^{\tilde{\Lambda}}$), with the latter representing an effective lower bound. 
%Fig.~2 shows the bound positron wavefunctions (also see Extended Data Table 1 for anisotropic polarizabilities).  %In addition to binding energies, solution of the Dyson equation also gives the fully-correlated positron-bound state wavefunctions (Dyson orbitals). These are shown in Fig.~2.
%Figure~2 shows the bound-state positron wavefunction densities. % for the polar molecules considered above (and also several non-polars to be considered below). 
%alongside comparison with reference values. 

 \renewcommand{\tablename}{}
\begin{table*}[h!!]
\footnotesize
\hrule
\vspace*{5pt}
\caption*{{\bf  \normalsize ~Table 1: Calculated positron-molecule binding energies (meV)}\label{table:be}}
\vspace*{-5ex}
\begin{center}
\begin{tabular}{l@{\hskip2pt}c@{\hskip2pt} c@{\hskip2pt}c@{\hskip8pt} c@{\hskip4pt} c@{\hskip4 pt}c@{\hskip4pt} c@{\hskip4pt} c@{\hskip8pt}c@{\hskip7pt} c@{\hskip4pt} c@{\hskip3pt} c@{\hskip3pt} c@{\hskip3pt}c@{\hskip3pt}}
\hline\\[-1.5ex]
&\multicolumn{3}{c}{}&&\multicolumn{4}{c}{Present many-body theory}  & & \multicolumn{4}{c}{Other calculations}\\
%\cmidrule(lr){6-9}\cmidrule(lr){10-12}\cmidrule(rr){13-17}
\cmidrule(rr){6-9}\cmidrule(rr){11-15}
			&$\mu$\,(D) &$\alpha$\,(\AA$^3$) &$I$\,(eV) & HF		& $\Sigma^{(2)}$ 		& $\Sigma^{GW}$ 	& $\Sigma^{GW+\Gamma}$ 	& $\Sigma^{GW+\Gamma+\Lambda}$\blue{$^{\dag}$}   	& Exp.\blue{$^{\ddag}$} & HF 	& CI 		& ECG\cite{bubin}  	& APMO &  \\
\hline
{\bf Polars} 	&&&					&		&				&		&			&				&							&							&		&					\\
LiH  			&5.9&3.50&8.3			&130 	& 434 			& 518 	&1291 		& 1106, 1038, {\bf 1060} 		& - 	&130\cite{Kurtz81}			&463\cite{STRASBURGER199649} 	& 1043 	&- 					\\
%HCN		&16.0&14.0& 2		&27			&7		&24		&23		&111			& 67				&62								& - 		&2		&35		& 38		& 8	\\%	&31-82\\
Acetonitrile	&3.9&4.24&12.6		&15		&120				& 109	& 301		& 210, 195, {\bf 207}		& $180\pm10$	&15\cite{Tachikawa11}	&136\cite{Tachikawa14}			& -		&65\cite{APMO2014}	\\
Propionitrile	&4.1&5.90&12.4		& 16		& 140			& 129	& 341		& 245,\,230,\,{\bf 243}		& $245\pm10$	&18\cite{Tachikawa11}	&164\cite{Tachikawa11} 		&- 		&- 					\\
Acetone		&2.9&5.75&10.2		& 3		& 67				& 69		& 215		& 143,\,135,\,{\bf 147}		& $174\pm10$	&- 					&96\cite{Tachikawa14}			&-		&36\cite{APMO2014}	\\ %1	\cite{Tachikawa03}
Propanal		&2.5&5.70&10.4		& 1		& 44				&45		&170			& 108, 100, {\bf 108}			&$118	\pm10$	&-					&58\cite{Tachikawa12}			&- 		& -					\\
Acetaldehyde	&2.7&4.12&10.6		& 2		& 35				&38		&135			& 86, 81, {\bf 89}			& $88\pm10$		&-					&55\cite{Tachikawa14}			&-		&16\cite{APMO2014}	\\
Formamide	&3.7&3.68&11.0		& 12		&105				&109		&255			& 186, 178, {\bf 189} 		& $\sim 200^\blue{\ast}$	&-		&-							&-		&-		\\
{\bf Nonpolars}&		&&			&		&				&		&			&				&		&					&							& 		&		\\
CS$_2$		&0&8.00&10.5			& $<0$	&$<0$		&$<0$	&171				& 68,\,46,\,{\bf 62}			&$75\pm10$		&-					&$<0$\cite{Koyanagi13}			&-		&-		\\
CSe$_2$		&0&10.7&9.7			& $<0$	&9 			&$<0$	&276				& 139,\,101,\,{\bf 131}			&-		&-					& 18\cite{Koyanagi13}			&-		&-		\\
Benzene		&0&9.85&9.5			&$<0$	&11				&	2	& 252		& 120,\,92,\,{\bf 116}		& 150	& -					& -							& -		& -		\\
\hline\\[-5ex]
\end{tabular}
\end{center}
\renewcommand{\tablename}{}
\caption*{\footnotesize % {\bf Many-body theory calculated positron binding energies and comparison with experiment and previous theory}. 
Dipole moment $\mu$ from [\citen{nistCCC}]; isotropic polarizabilities $\alpha$ and ionization energies $I$ calculated at the $GW$ level (see Extended Data Table 1 for anisotropic polarizabilities). 
Binding energy calculations are presented in the Hartree-Fock,  $\Sigma^{(2)}$ (bare-polarisation) and $GW$@BSE (bare-polarisation plus screening and electron-hole corrections) approximations, 
%various levels of the $GW$ approximation:
%(see Fig.~1): 
%$\Sigma^{(2)}$, 
% (corresponding to the bare polarization propagator); 
%RPA, 
%% (random phase approximation); 
%TDHF,
%% (time-dependent Hartree Fock approximation); and 
%BSE,
% (Bethe-Salpeter equation), 
and additionally including virtual-positronium formation $\Sigma^{GW+\Gamma}$, 
and the positron-hole ladder contribution $\Sigma^{GW+\Gamma+\Lambda}$:  
\blue{$^{\dag}$}for the latter, the first (second) number is that using bare (dressed) Coulomb interactions in the $\Gamma$ and $\Lambda$ blocks, and the third (our most sophisticated calculation, in bold) additionally uses $GW$ energies in the diagrams. Their difference gives a measure of the theoretical uncertainty; \blue{$^\ddag$}Experiment from \cite{RevModPhys.82.2557,Danielson10,Danielson12}; \blue{$^{\ast}$}except that for formamide, which is preliminary and unpublished\cite{pJames}. 
Other calculations: ECG (explicitly correlated Gaussian), CI (configuration interaction), and any-particle-molecular orbital (APMO, `REN-PP3').
Also see Fig.~2 for a graphical comparison of experiment with present and previous theory. % and Extended Data Table 2 for more .
%
%$^1$ H. A. Kurtz and K.  D. Jordan, J. Chem. Phys. 75, 1876 (1981),\\
%$^2$ K.  Strasburger, Chem. Phys.  Lett. 253, 49-52,(1996), \\
%$^3$ S. Bubin and L. Adamowicz,  J.  Chem.  Phys. 120,6051 (2004),\\
%$^4$ M. Tachikawa, Y. Kita and R. J. Buenker, Phys. Chem. Chem. Phys. 13, 2701 (2011),\\
%$^5$ M. Tachikawa, J. Phys. Conf. Ser. 488, 012053 (2014),\\
%$^6$ J. Romero et.al., J. Chem.Phys. 141 (2014), \\
%$^7$ M. Tachikawa, Y. Kita and R. J. Buenker, New J. Phys. 14, 035004 (2012),\\
%$^8$ K. Koyanagi et al., Phys. Chem. Chem. Phys. 15, 16208 (2013)
} %and additionally including the positron hole ladder series￼}
%The suggested best results are the HF + GW@BSE + ￼.... 
%The error on the experimental values is ~?meV... (can we mention the systematic error? Show calculated polarisaibilities: why is our CS2 polarizability ~8 instead of 10... multiplying 61*10/8 gets us to 75
\vspace*{-5pt}\hrule
\label{default}
\end{table*}%
\renewcommand{\figurename}{{\bf \footnotesize }}
\begin{figure*}[h!!]
\vspace*{-4ex}
\centering
\includegraphics[width=0.85\textwidth]{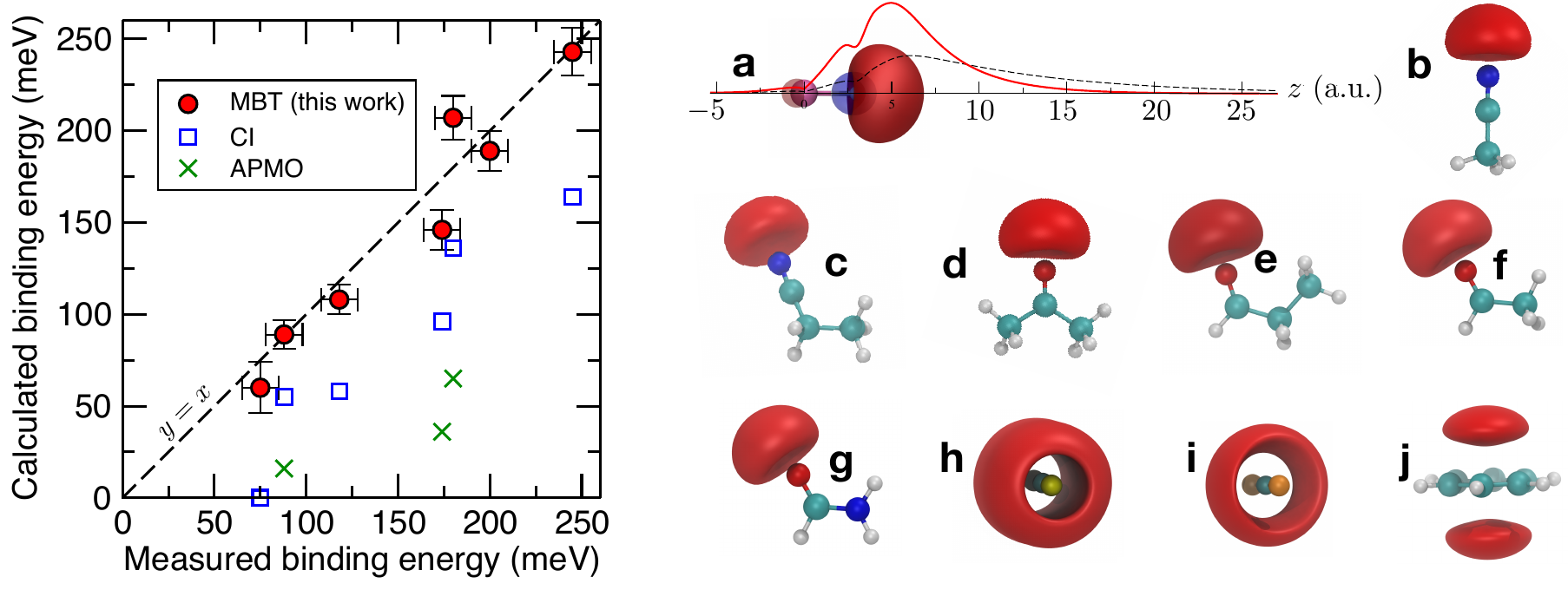}
%\includegraphics[width=0.98\textwidth]{wavefunctions-small}
%\includegraphics[width=1\textwidth]{wavefunctions-contactdensities}
%\includegraphics[width=1\textwidth]{contactdensities-small}
%\begin{figure}
%\includegraphics[width=1\textwidth]{CDwireframe}
%\end{figure}
%\includegraphics[width=0.5\textwidth]{CS2xz2d} \includegraphics[width=0.5\textwidth]{CS2xz2d}
\vspace*{-1ex}
\caption*{\footnotesize {\bf Fig.~2. Positron-molecule binding energies and bound-state Dyson wavefunction densities.}  
Graph shows the comparison of the present many-body calculations (red circles) with experiment. Also shown are the CI and APMO calculations (blue square and crosses, respectively). Positron wavefunction densities:
{\bf a}, LiH, with Li atom at origin and H at $\sim$3 a.u.~along the molecular axis, showing the positron wavefunction density isosurface at 70\% of the maximum (red lobe), the electron HOMO wavefunction density isosurface (blue lobe is negative region at 40\% of maximum, and brown is positive region at 10\% of the maximum).
%To emphasise the diffuseness of the positron wavefunctions, 
Also shown is the positron wavefunction calculated along the molecular axis in the static HF approximation (black curve) and at the $\Sigma^{GW+\Gamma+\Lambda}$ level of many-body theory (red curve).
%Other molecules:
{\bf b}, acetonitrile;
{\bf c}, propionitrile; 
{\bf d}, acetone;
{\bf e}, propanal; 
{\bf f}, acetaldehyde;
and
{\bf g}, formamide;
show the positron wavefunction density isosurface at 80\% of the maximum.
For the nonpolars:
{\bf h}, CS$_2$;
{\bf i}, CSe$_2$; and  
{\bf j}, benzene, the isosurfaces are at 90\% of the maximum. %positron wavefunction density. 
%Atoms are coloured as follows: carbon, green; hydrogen, grey; nitrogen, blue; oxygen, red; sulphur, yellow; selenium, orange.
}\label{fig:wavefunctions}
\end{figure*}

%\smallskip
%\smallskip
\vspace*{-1.5ex}
\noindent{\bf \emph{Benchmarking and general trends}}.---We benchmark our approach against a highly-accurate explicitly correlated gaussian (ECG) calculation ($\varepsilon_b=1043$ meV)\cite{bubin} for the strongly polar molecule LiH. %exists against which we benchmark our approach. %providing a benchmark for our approach. 
The results demonstrate the general trends seen in all the molecules considered.
The HF binding energy ($\varepsilon_b=130$\,meV) is severely deficient.  %neglects correlations and  % is in perfect agreement with previous the HF calculation\cite{Kurtz81}, but 
%severely underestimates the true binding. % due to the absence of correlations. 
%Including the $GW$ self-energy at the successive levels of sophistication  
%(Also see Extended Data Table 2 for corresponding calculated isotropic molecular polarizabilities $\alpha$). 
Including the bare polarization attraction $\Sigma^{(2)}$ significantly increases the binding energy (to $\varepsilon_b=434$\,meV).
The addition of short-range screening corrections %($GW@$RPA) 
reduces the polarizability and binding energy (to $\varepsilon_b=336$\,meV, see Extended Data Table 2), but this is 
compensated by the inclusion of the electron-hole attractions 
%(GW$@TDHF: $\varepsilon_b=542$\,meV; and $GW@$BSE: $\varepsilon_b=518$\,meV). 
%(GW$@TDHF: $\varepsilon_b=542$\,meV; and 
($\Sigma^{GW}$: $\varepsilon_b=518$\,meV). 
%For all the molecules considered, comparing the binding energy calculated using the bare polarization diagram $GW@\Sigma^{(2)}$ to that using the more sophisticated (and computationally demanding) $GW$@BSE  shows that the effect of screening of the electron-positron Coulomb interaction by molecular electrons, and electron-hole corrections act to effectively cancel one another. 
%Compared to the accurate ECG calculation ($\varepsilon_b=1043$\,meV), the polarization potential ($GW$@BSE) 
This is still, however, less than half of the ECG result. The previous CI calculation\cite{STRASBURGER199649} is similarly deficient. 
Strikingly, however, including the virtual-positronium formation correlation potential ($\Sigma^{GW+\Gamma}$) strongly enhances the binding, more than doubling it
% the binding energy from the $GW$@BSE result for all the molecules considered 
 (to $\varepsilon_b=1291$\,meV). 
Including the positron-hole ladder ($\Sigma^{GW+\Gamma+\Lambda}$) slightly reduces binding (to $\varepsilon_b=1106$ meV); using screened interactions in the ladders 
%($\Sigma^{GW+\tilde{\Gamma}+\tilde{\Lambda}}$) 
reduces it slightly further ($\varepsilon_b=1038$\,meV); and additionally using the dressed energies in the diagram construction gives $\varepsilon_b=1060$\,meV, agreeing with the ECG result to $\sim$1\%. 
%For all the
As for all the  polar molecules, the \emph{maximum} of the positron wavefunction density (Fig.~2) in LiH is highly localized at the negative end of the molecule, 
%One should not, however, be deceived: 
%Overall, however, 
%The positron bound-state wavefunction 
but overall the wavefunction is 
%nevertheless 
quite diffuse, asymptotically taking the form $\psi\sim e^{-\kappa r}$ where $\kappa=\sqrt{2\varepsilon_b}$.
We also calculate the positron Dyson wavefunction renormalization constants $a$ (see `Methods Eqn.~7 and Extended Data Table 2). These represent the contribution of the ‘positron plus molecule in the ground state’ component to the positron-molecule bound state. Their closeness to unity suggests the picture of a positron bound to the neutral molecule (rather than a Ps atom orbiting a molecular cation) \cite{Mitroy_2002}. 

\smallskip
%\smallskip
\noindent{\bf\emph{Comparison with experiment and previous theory}}.---%We now consider the six molecules for which \emph{ab initio} theory and measurements both exist. % we see a similar picture.  
The best prior agreement between theory and experiment \emph{for any molecule} was for acetonitrile
%the CI calculation ($\varepsilon_b$= 136\,meV\cite{Tachikawa14}) agreeing with experiment ($\varepsilon_b$= 180\,meV\cite{Danielson12}) to 
($\gtrsim 25\%$). % accuracy. 
Considering the polar molecules first (Table 1 and Fig.~2), we immediately see that our full many-body theory ($\Sigma^{GW+\Gamma+\Lambda}$) is much superior, giving near exact agreement ($\lesssim1$\% level) with experiment for propionitrile, propanal, acetaldehyde and formamide,
% for this molecule the full many-body theory result ($\Sigma^{GW+\Gamma+\Lambda}$: $\varepsilon_b$= 195--210\,meV) 
%and ($\Sigma^{GW+\tilde{\Gamma}+\tilde{\Lambda}}$: $\varepsilon_b$= 195\,meV) 
and within 10\% for acetonitrile and acetone (including the experimental error).  
(Overall we find excellent convergence in our calculation: see `Methods' and Extended Data Fig.~2). 
For all the polar molecules, the HF and bare ($\Sigma^{(2)})$ and dressed ($GW$) polarization results significantly underestimate binding.
The effect of virtual-positronium is crucial: it enhances the binding energy by a factor of $\sim$2 and is essential to bring theory into agreement with experiment. 
We note that the previous configuration interaction and any-particle-molecular orbital (renormalized PP3 “REN-PP3”, which employs a diagonal approximation and does not explicitly account for virtual-positronium formation) calculations are severely deficient. The ECG approach is not easily scalable to these molecules\cite{bubin}.
%(see Fig.~2 for graphical comparison). %the latter , being a factor of $\sim 3$ deficient.\\ 
%Our calculated polarizability and ionization energies are in excellent agreement with reference values (see Extended Table 2), and we observe good basis-set convergence (e.g., see Extended Data Fig.~1). 

For the nonpolars, binding is exclusively enabled by correlations. 
For CS$_2$ a considerable binding energy of $75$ meV has been measured, whilst the CI calculation failed to predict binding\cite{Koyanagi13}. 
We find that polarization correlations ($GW$) alone are insufficient to support binding. Spectacularly, however, including the virtual-positronium contribution results in significant binding: 
our $\Sigma^{GW+\Gamma+\Lambda}$ result of $\varepsilon_b=46$--68 meV is close to experiment: the range is larger than in the calculations for the polars as the delocalization of the positron wavefunction (Fig.~2 h--j) makes accurately describing virtual-positronium more demanding.
For CSe$_2$ and benzene, in contrast to the molecules already considered, we have not performed any optimization of the bases (due to the delocalised positron wavefunction, these molecules require computational resources beyond our disposal)
% (e.g., have used only 3 additional ghost centres for benzene and 8 centres for CSe$_2$). 
%Our calculations for these two molecules 
and our calculated $\varepsilon_b$ should be considered as lower bounds. 
Nevertheless, the results further elucidate the essential role of virtual-positronium formation in enabling (significant) binding, and the positron wavefunctions also provide fundamental insight that may prove instructive to refine \emph{ab initio} and model calculations (see below).

\noindent{\bf\emph{Prediction for formamide}}.---For this, the archetypal molecule for the investigation of protein and peptide chemistry, we are unaware of any prior calculation. We predict binding ($\varepsilon_b\sim189$ meV). Preliminary experiments see evidence of $\varepsilon_b\sim$200\,meV, although a final value has yet to be determined \cite{pJames}.

\begin{figure*}[bt!!]
\centering
\includegraphics[width=1\textwidth]{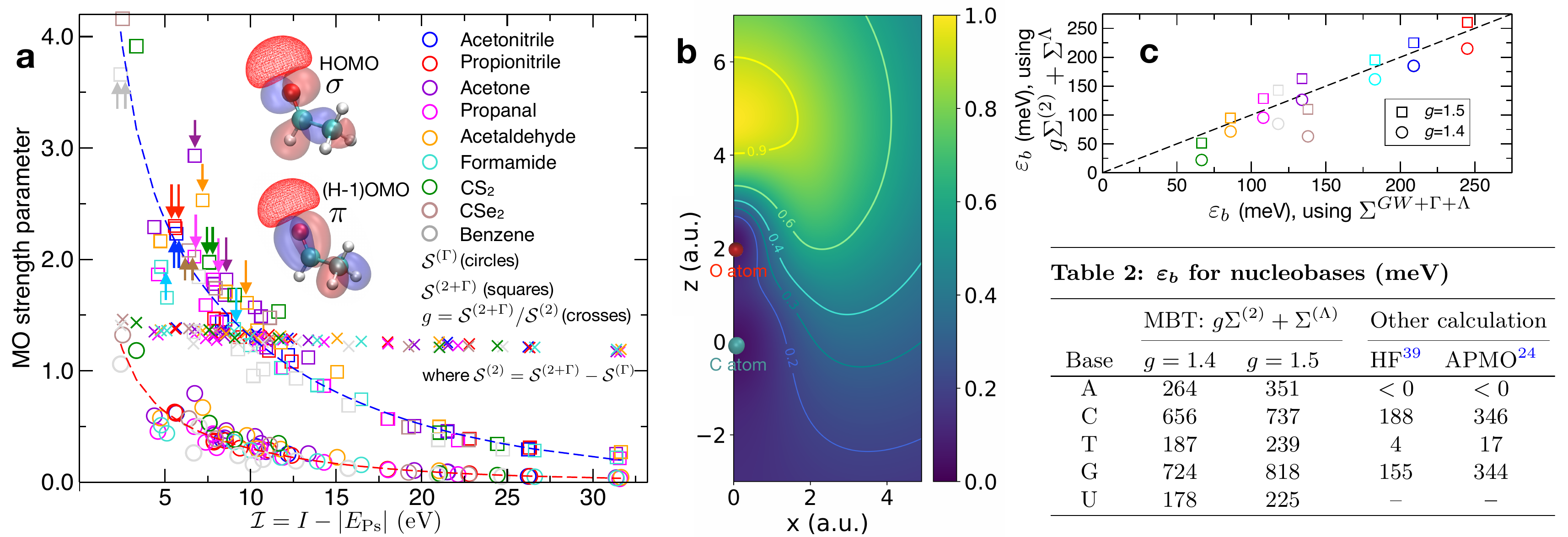}
%\includegraphics[width=1\textwidth]{contactdensities-small}
%\begin{figure}
%\includegraphics[width=1\textwidth]{CDwireframe}
%\end{figure}
%\includegraphics[width=0.5\textwidth]{CS2xz2d} \includegraphics[width=0.5\textwidth]{CS2xz2d}
\caption*{\footnotesize {\bf Fig.~3. Molecular orbital contributions to binding, and scaling formula for large molecules.} {\bf a}, MO contribution to the dimensionless strength 
of the virtual-positronium formation correlation potential $\mathcal{S}^{(\Gamma)}$ (circles) and including bare polarization $\mathcal{S}^{(2+\Gamma)}$ (squares) and the ratio $g\equiv\mathcal{S}^{(2+\Gamma)}/\mathcal{S}^{(2)}$ (crosses, where $\mathcal{S}^{(2)}=\mathcal{S}^{(2+\Gamma)}-\mathcal{S}^{(\Gamma)}$) against $\mathcal{I}=I-|E_{\rm Ps}|$, where $I$ is the MO ionization energy and  $E_{\rm Ps}=-6.8$ eV is the ground state energy of Ps. Arrows on $\mathcal{S}^{(2+\Gamma)}$ mark $\pi$ orbitals with $I<15$ eV, and inset MO plots show HOMO ($\sigma$ type) and next HOMO ($\pi$ type) % with relatively larger contribution to the strength) 
in acetaldehyde (solid red and blue: positive and negative regions of electronic MO; red wireframe: positron density at 85\% of maximum). 
Dashed lines show the fits  
% 2+ GAMMA
$\mathcal{S}\approx ae^{-b\mathcal{I}}+c\mathcal{I}^{-d}$ %$\mathcal{S}\approx ae^{-b\mathcal{I}^c}+d\mathcal{I}^{-e}$
%\red{FIX FIT... ONE THING DIDNT WORK}
with
$
a=0.67\,(2.57); 
b=0.121\,(0.092); 
c=2.51\,(6.64);
d=1.38\,(1.37)
$
for $\mathcal{S}^{\Gamma}$ ($\mathcal{S}^{2+\Gamma}$).
{\bf b}, the positron wavefunction density in acetaldehyde (in the plane containing the CO bond perpendicular to the CCO plane), which protrudes along the $\pi$ bond. 
{\bf c}, comparison of binding energies for the molecules in {\bf a} calculated using $\Sigma=\Sigma^{GW+\Gamma+\Lambda}$ and accounting for (the computationally demanding to calculate) $\Sigma^{\Gamma}$ via $\Sigma=g\Sigma^{(2)}$+$\Sigma^{\Lambda}$, for $g=1.4$ (circles) and 1.5 (squares) (see text). {\bf Table 2}, $\varepsilon_b$ in nucleobases, calculated with $\Sigma\approx g\Sigma^{(2)}$+$\Sigma^{\Lambda}$ for $g=1.4$ and $g=1.5$.
%\red{FIX ARROWS FOR NONPOALRS}.  
%\red{1. MBT acronym defined?  3. Colours in (a) for non-polars}%: the HOMO (b) contributes a smaller strength compared to the $\pi$ (H-1)OMO (c).   %HOMO, HOMO-1, HOMO-2 and HOMO-3 
%{\bf e} and {\bf f}, the two HOMOs of acetone, and {\bf i}, positron density in acetone in the plane containing the oxygen atom perpendicular to the CC bonds. 
}\label{fig:cds}
\end{figure*}
%\red{Dipole/polarizability/Anisotropy? Strength? }
%, and estimating {$\boldsymbol\varepsilon_b$} in large molecules: nucleobases
\begin{comment}
\red{Whilst In Table 1 we quote the isotropic polarizability and ionization energy of the molecules considered, our results indicate that the mechanisms of positron binding cannot be described in terms of such coarse properties. 
The interactions are inherently anisotropic, and individual molecular orbitals (of different symmetry and ionization energies) contribute significantly to the correlation potential. }
%The anisotropic nature of the interaction is evidenced through the calculation of t
\end{comment}
\smallskip
\noindent{\bf{\emph{Molecular orbital contributions to binding: anisotropy and strength of correlations.}}}---At the static HF level, we find $\varepsilon_b$ to be (monotonically and non-linearly) related to the permanent dipole moment (expected from the dipole-potential model\cite{Gribakin2015}).
Ultimately the correlation potential is anisotropic (see Extended Data Table 1 for calculated anisotropic polarizabilities), and depends non-linearly on the polarizabilities and ionization energies of the individual MOs. Moreover, the binding energy depends non-linearly on the correlation potential (e.g., see Extended Data Fig.~3).
The ordering of $\varepsilon_b$ with respect to dipole moment persists to the full $\Sigma^{(2+\Gamma+\Lambda)}$ calculation, with the exception of acetaldehyde and propanal, and we note that % at the $\Sigma^{(2)}$ level $\varepsilon_b$ is larger in the latter, 
for acetone correlations considerably enhance $\varepsilon_b$.
It is instructive to consider the dimensionless quantity $\mathcal{S} = - \sum_{\nu>0} \varepsilon_{\nu}^{-1}\langle\nu|\Sigma|\nu\rangle $ \cite{Dzuba:1994} (where the sum is over excited HF positron basis states of energy $\varepsilon_{\nu}$, see `Methods'), which gives an effective measure of the strength of the correlation potential $\Sigma$. 
The magnitudes of the strength of $\Sigma^{(2)}$, $\mathcal{S}^{(2)}$ ranges from 4--15 (see Extended Data Table 2), and follows the ordering of the isotropic polarizability, with the exception of acetone and propanal (acetone has a larger polarizability and smaller ionization energy than propanal), and benzene and CSe$_2$ (owing to benzene's $\pi$ bonds, see below).
% and the MO below the HOMO contributes most (see below). 
This suggests that (the short range contributions to) $\Sigma^{(2)}$ cannot be parametrized solely by the polarizability.
%$\Sigma^{(2)}$ is larger for propanal than acetaldehyde. 
Similarly, the magnitudes of $\mathcal{S}^{(\Gamma)}$ (ranging from 2--5) do not strictly follow the ordering of the ionization energies.
%: a smaller difference in the ionization energy and the positronium-formation energy leads to greater probability of the electron tunnelling to the positron.
%$\varepsilon_{\rm Ps} = I + E_{\rm 1s}({\rm Ps}) = I - 6.8$ eV, where $I$ is the ionization potential and $E_{\rm 1s}$(Ps) is the ground-state energy of Ps). 
%\red{The fact that numerous molecular orbitals contribute to binding can be understood by consideration of the , which is an effective measure of the correlation potential. }
To illuminate this, note that at the bare-polarization, $\Sigma^{(2)}$, and polarization plus virtual-positronium formation approximations, $\Sigma^{(2+\Gamma)}=\Sigma^{(2)}+\Sigma^{(\Gamma)}$, we can delineate the contribution of individual MOs to positron binding. 
%\red{The strengths\dots}
% viz., $\mathcal{S}=\sum_{n} \mathcal{S}_{n}=\sum_n S_n^{(2)}+\mathcal{S}_n^{(\Gamma)}$, where $n$ labels the occupied electronic MO. 
Fig.~3\,a shows the partial $\mathcal{S}^{(\Gamma)}$ and $\mathcal{S}^{(2+\Gamma)}$ for individual occupied MOs against their respective ionization energies, and the ratio $g\equiv \mathcal{S}^{(2+\Gamma)}/\mathcal{S}^{(2)}$, where $\mathcal{S}^{(2)}=\mathcal{S}^{(2+\Gamma)}-\mathcal{S}^{(\Gamma)}$. %Both 
%First, we note that relative strength of the bare-polarization term and the virtual-positronium formation one are similar. 
%Secondly, we note that 
%In propanal, the (H-3)OMO of $\pi$ type contributes more strongly than the (H-2)OMO.
%the HOMO does not necessarily contribute most to either $\Sigma^{(\Gamma)}$ or $\Sigma^{(2+\Gamma)}$: 
Both $\mathcal{S}^{(\Gamma)}$ and $\mathcal{S}^{(2+\Gamma)}$ decrease from the Ps-formation threshold to higher ionization energies: it is more difficult to perturb more tightly bound electrons. %, either to contribute to polarization or virtual-positronium formation. 
However, the decrease is not monotonic: we see that despite having larger ionization energies, $\pi$-type electronic MOs below the HOMO can contribute significantly more than a $\sigma$-type HOMO to $\mathcal{S}^{(\Gamma)}$ and $\mathcal{S}^{(2+\Gamma)}$, e.g., in acetone, propanal, and acetaldehyde, the strength of the $\pi$-type (H-1)OMO is larger than the $\sigma$-type HOMO, and in propanal, the (H-3)OMO of $\pi$ type contributes more strongly than the (H-2)OMO etc. It was previously speculated in Ref.~[\citen{Danielson09}] that $\pi$ bonds were important due to the ability of the positron to more easily access electron density that is delocalized from (repulsive) nuclei. This is borne out by our calculations, and we see that in Fig.~3\,b considerable positron density protrudes into the region of the $\pi$ bond. Acetonitrile and propionitrile have a doubly-degenerate $\pi$ HOMO of large strength. For acetonitrile this results in a larger strength parameter than formamide.

%%paints a 
%This complex picture makes 
%
%searching for a simple \emph{universal} 
% may be somewhat futile,
%%%from which obtaining a simple universal scaling formula based on coarse molecular properties  \cite{Danielson09,Danielson12,MLbind} appears challenging, likely 
%%%that 
%
%Whilst attempts at scaling formula have Coarse molecular properties, e.g., $\mu$, $\alpha$ and (first) ionisation energy alone will be poor \emph{universal} predictors of binding energies.
% predictor: relative contribution and symmetries of MOs}%\\[-2ex] %in 

% the strong correlational enhancement to the dipole-induced static interaction, and overall the interactions are anisotropic. non-linear and ansiotropic interactionscannot be described solely by coarse molecular properties (e.g., . 

% and highlights that binding \red{depends on a number of interrelated properties}. 
% \red{coarse molecular properties, e.g., $\mu$, $\alpha$ Ionisation energy alone a poor predictor: relative contribution and symmetries of MOs}%\\[-2ex] %in the $\pi$-type MOs. 
%Examples include acetaldehyde, where the $\pi$-type MO below the HOMO (shown in Fig.~(3) (c)) has a considerably stronger contribution than the HOMO (Fig.~3(b)), and acetone (Fig.~3 (f)--(i)).
%Our work thus gives quantitative verification of the importance of $\pi$ bonds in positron binding, }].\\[-2ex] 

\smallskip
\smallskip
\noindent{\bf{\emph{Predicting positron binding in larger molecules: nucleobases as an example.}}}---The ratio $g\equiv\mathcal{S}^{(2+\Gamma)}/\mathcal{S}^{(2)}$ is weakly dependent on the ionization energy, with a value of $\sim$ 1.4--1.5 for the HOMOs ($I\sim10$ eV). %$ (this is also true for the non polars, which are not shown in the figure). 
%This, and the knowledge that short-range corrections to the polarization effectively compensate eachother, 
We propose that binding energies of large molecules (e.g.,  15--100 atoms, for which a converged calculation of the virtual-positronium diagram Fig.~2\,c is too computationally demanding) can be calculated by approximating $\Sigma\approx g\Sigma^{(2)}+\Sigma^{\Lambda}$. As well as accounting for virtual-positronium formation, this model potential reflects the anisotropy of the true interactions.  %i.e., accounting for the virtual-positronium formation contribution via $g$. 
For the molecules considered in Table 1, this works well (see Fig.~3 c and also Extended Data Fig.~3). 
Using this approximation, we calculate
the positron binding energy in the five nucleobases (Table 2). Our results are larger than the previous APMO calculations, mirroring the results for the molecules in Table 1. We predict binding in adenine.

\begin{figure}[t!!]
\centering
%\includegraphics*[width=0.5\textwidth]{CDsqrteb}
%\includegraphics*[width=0.98\textwidth]{cdi1}
%%\includegraphics*[width=0.48\textwidth]{CDvsI}
%%\includegraphics*[width=0.48\textwidth]{CDsqrteb_edit}
%\caption*{\footnotesize{{\bf Extended Data Figure 4. Calculated electron-positron contact density}. {\bf a}, contact density  for individual electronic MOs as a function of their ionisation energy, calculated  including vertex enhancement factors and renormalization coefficients (see `Methods' Eqns.~6 and 7). Red dashed line: positronium ground state energy at $|E_{\rm Ps}|=6.8$ eV. Grey line: $\delta_{ep}=0.008/(I-|E_{\rm Ps}|)$ (for a guide). Also see Extended Data Table 3 and Extended Data Fig.~5. 
%{\bf b}, Contact density calculated at the HF (circles) and  various levels of many-body theory (diamonds: $GW$@BSE; squares: $GW$@BSE+$\Gamma$+$\Lambda$) against the square root of the binding energy.}\\[4ex]
\includegraphics[width=1\textwidth]{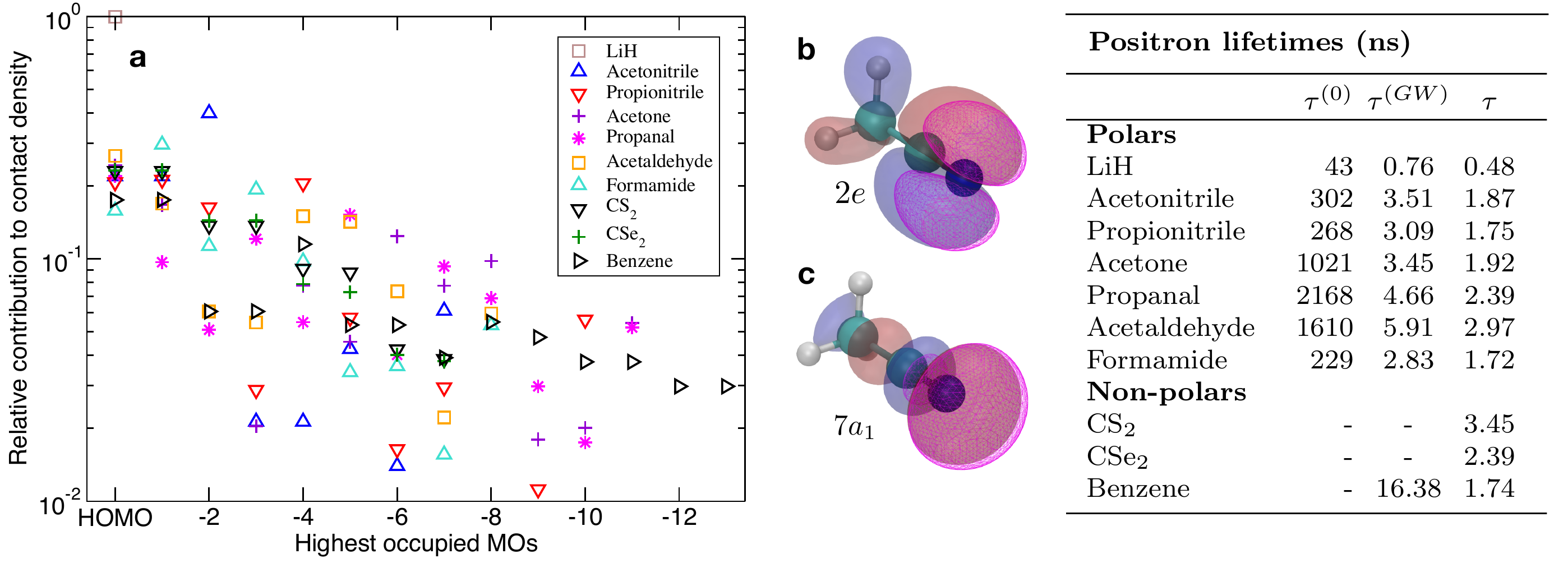} 
\caption*{\footnotesize {\bf Fig.~4. Calculated electron-positron contact densities and positron lifetimes with respect to annihilation.}  
{\bf a}, fractional contribution of individual MOs to the total electron-positron contact density (Eqn.~6 in `Methods).
%with x-axis counting down from the HOMO (highest-occupied molecular orbital) 
{\bf b} and {\bf c}, the electron-positron contact density (magenta) at the $\Sigma^{GW+\Gamma+\Lambda}$ level for the HOMO and (H-1)OMO
 in acetonitrile (blue and brown show positive and negative electron wavefunction regions, respectively), c.f., Fig.~2 (b) (positron density).
{\bf Table: Positron lifetimes with respect to annihilation}. $\tau^{(0)}$: lifetime calculated in the Hartree-Fock independent particle approximation excluding the vertex enhancement factors and using a positron wavefunction normalised to unity; $\tau^{GW}$ and $\tau$: lifetime calculated using the Dyson positron wavefunction at the $\Sigma^{GW}$ and $\Sigma^{GW+\Gamma+\Lambda}$ levels including vertex enhancement factors and renormalization constants (Eqns.~6 and 7 of `Methods'). 
}\label{fig:cds}
%\label{fig:acetocon}}
\end{figure}

\smallskip
\smallskip
\noindent{\bf{\emph{Positron lifetimes.}}}
Correlations and the relative MO symmetries also dictate the lifetime $\tau\,{\rm [ns]} \approx 0.02\,\delta_{ep}^{-1}\,{[\rm a.u.]}$ of the bound-state positron against annihilation, where $\delta_{ep}$ is the electron-positron contact density [see `Methods' Eqn.~(6) and Extended Data Table 3 for values of $\delta_{\rm ep}$]. 
We also calculate the annihilation lifetime of the bound positron (see Figs.~4, `Methods' and Extended Data Fig.~4), finding that the correlations reduce it by  two orders of magnitude to $\tau \sim 1$ ns,  
 %due to the increased electron-positron overlap, and moreover, that this overlap is heavily 
and find that the contribution of electron MOs to annihilation depends strongly on their  symmetry relative to that of the positron MO, with the HOMO not necessarily dominating, e.g., in acetonitrile and formamide. % (\red{see Extended Data Fig.~)}. 
\section*{Conclusions and future perspectives.}
%The many-body theory developed here to open fundamental problem of positron binding to molecules.
%Whilst measurements for over 80 molecules have been made over the past few decades, an \emph{ab initio} description has remained elusive. The previous quantum chemistry approaches were found to be severely deficient, agreeing at best to only within 25\% of experiment (for acetonitrile), and failing to predict binding in non-polar molecules.
%
%We developed a many-body theory of positron interactions with polyatomic molecules and 
%applied it, via a state-of-the-art computational implementation, to the six polar organic and non-polar molecules for which theory and experimental results exist (in addition to LiH, formamide, CSe$_2$, and benzene).

Many-body theory has uncovered the mechanisms of positron binding to molecules. 
Binding is governed by the interplay of the static (dipole) interaction with the strong (and non-local) correlation potential,
to which numerous MOs contribute significantly. The interactions are inherently anisotropic, and the binding energy depends non-linearly on the correlation potential. 
%long-range polarization, short-range screening, electron-hole attraction and virtual-positronium formation effects, 
%and a non-linear dependence of binding on the correlation potential. 
%Delineating the effects of distinct correlations and the contribution of individual molecular orbitals, we uncovered 
In particular, we uncovered the key role of
virtual-positronium formation in significantly enhancing and enabling binding, and quantified the importance of $\pi$ bonds. 
%in polar molecules, and is essential to support binding in nonpolars. 
Overall, 
the \emph{ab initio} approach, which takes proper account of the distinct correlations and the ansiotropy of the interactions gives
%we obtain 
binding energies in the best agreement with experiment to date (highlighting the severe deficiency of previous quantum chemistry calculations). We also predicted binding in formamide and nucleobases.

The present work directly supports resonant annihilation experiments and related model theory that requires positron binding energies as parameters \cite{RevModPhys.82.2557}. 
Moreover, the many-body theory approach can be naturally extended to describe positron scattering and (non-resonant) annihilation rates and $\gamma$ spectra in molecules and condensed matter.
%with numerous important implications (see e.g., \cite{Danielson:2015,fajans:2020).  Calculations of relative probabilities of annihilation on individual MOs in molecules may 
Calculations of $\gamma$ spectra can provide insight on energy deposition and molecular fragmentation \cite{ionprod,posfrag,nrgdepo,frag2}, clusters and cluster surfaces\cite{fajans:2020}, and the determination of electron-momentum densities from annihilation $\gamma$ spectra measurements \cite{DGG_molgamma}. 
As well as novel positron-based molecular spectroscopy, more generally, such predictive capability could provide fundamental insight required to develop positron traps, accumulators and high-energy-resolution beams\cite{Danielson:2015,fajans:2020} (for e.g., shorter temporal pulses in lifetime spectroscopy, colder positrons for antihydrogen formation, and higher densities for Ps BEC production\cite{Schippers_2019}), to develop next-generation positron emission tomography (including understanding positron emitting tracers and positron-induced DNA damage, and new spectroscopic PET\cite{Schippers_2019}), and proper interpretation of positron-based materials science diagnostics %techniques that probe defects and time-dependent surface processes of industrially important materials
\cite{RMPpossolids2013,hugreview}. 
%which would e.g., improve positron-based molecular spectroscopy, drive efforts to produce Ps-BEC and 
%The develop
Furthermore, the formation and interrogation of antihydrogen\cite{Baker:2021} gives promise for more complicated anti-atoms/molecules: study of their structure, interactions and chemistry will require theoretical understanding.
%Moreover, with the promise of the production of more complicated antimatter atoms and molecules, developments of antimatter-matter structure calculations such as those presented here will be essential.
Finally, the many-body formalism provides a natural foundation to develop \emph{ab initio} descriptions of positron-induced (excited-state) molecular processes, such as positron-induced interatomic Coulomb decay or positron capture\cite{icdenv}, charge migration and luminescence \cite{PhysRevLett.120.147401}, and (whilst extremely challenging) the inclusion of phonons and photons, which may enable \emph{ab initio} description of vibrational Feshbach resonant annihilation spectra via coupling of the nuclear and electronic degrees of freedom, 
%\cite{Brand:1999}, 
or modelling of positron pump-probe experiments\cite{RMPpossolids2013}.

%\newpage
\section*{Methods}

%\red{NEED TO ADD THE polarization PROPAGATOR AND VIRTUAL-PS HERE}

\small{
\subsubsection*{Solving the Dyson equation for the positron binding energy and wavefunctions using a Gaussian basis.}
We calculate positron-molecule binding energies $\varepsilon$ and quasiparticle wavefunction $\psi_{\varepsilon}$ by solving the Dyson equation, main text Eqn.~(1). We take the zeroth-order Hamiltonian $H^{(0)}$ to be that of the positron in the Hartree-Fock field of the frozen-target $N$-electron ground-state molecule. 
The self-energy diagrams thus begin at second order in the Coulomb interaction. Rather than computing the self energy $\Sigma({\bf r,r}')$ in the coordinate basis, it is more convenient to work with its matrix elements in the Hartree-Fock basis.
Specifically, we expand the electron (-) and positron (+) Hartree-Fock MOs $\varphi_{a}^{\pm}(\bm{r})$ in distinct Gaussian basis sets 
as $\varphi_{a}^{\pm}(\bm{r})=\sum_{A}^{N_c^{\pm}}\sum_{k=1}^{N_A^{\pm}} C_{a Ak}^{\pm} \chi^{\pm}_{A_k}(\bm{r})$, 
where $A$ labels the $N_c^{\pm}$ basis centres, $k$ labels the $N_A^{\pm}$ different Gaussians on centre $A$, each taken to be of Cartesian type with angular momentum $l^x+l^y+l^z$, viz.,
$\chi_{A_k}(\boldsymbol{r}) = \mathcal{N}_{A_k}(x-x_A)^{l^x_{Ak}}(y-y_A)^{l^y_{Ak}}(z-z_A)^{l^z_{Ak}} \exp\{{-\zeta_{Ak} |{\bf r-r}_A|^2}\}$, where $\mathcal{N}_{A_k}$ is a normalisation constant, and $C$ are the expansion coefficients to be determined (see below). 

For both electrons and positrons, we use the diffuse-function-augmented correlation-consistent polarised aug-cc-pVXZ (X=T or Q) Dunning basis sets centred on all atomic nuclei of the molecule, which enables accurate determination of the electronic structure including cusps\cite{Kato} and expulsion of the positron density from the nuclei. 
To capture the long-range correlation effects, for the positron we also additionally include at least one large even-tempered set at the molecular centre or region of maximum positron density of the form $Ns(N-1)p(N-2)d(N-3)f(N-4)g$ with $N\sim10-15$ (where it should be understood that the full degenerate set of non-zero angular momentum functions is used) and exponents $\zeta_{A_k} = \zeta_{A_1}\beta^{k-1}$, $k=1,\dots, N$, for each angular momentum type, where $\zeta_{A_1}>0$ and $\beta>1$ are parameters. The value of $\zeta_{A_1}$ is important as the bound positron wavefunction behaves asymptotically as $\psi\propto e^{-\kappa r}$, where $\kappa=\sqrt{2\varepsilon_b}$. Thus, to ensure that the expansion describes the wavefunction well at $r\sim 1/\kappa$, i.e., that the broadest Gaussian covers the extent of the positron-wavefunction, one must have $\zeta_{A_1}\lesssim \kappa^2=2\varepsilon_b$.
In practice we performed binding energy calculations for a range of $\zeta_{A_1}$ and $\beta$ for each molecule, finding that there are broad ranges of stability.  %which are related to the value of $\varepsilon_b$, therefore some prior estimate of $\varepsilon_b$ can be beneficial in these optimizations.  
The optimal $\zeta_{A_1}$ was typically found to be in the range of $10^{-4}-10^{-3}$ for $s$- and $p$-type Gaussians and $10^{-3}-10^{-2}$ for $d$- and $f$-type Gaussians, whilst $g$-type Gaussian exponents usually had $\zeta_{A_1}=10^{-1}$ (atomic units are assumed throughout unless otherwise specified). The optimal $\beta$ ranges from 2.2 to 3.0 depending on the number of functions $N$ in a given shell.
Finally, to improve the description of the virtual-Ps formation process, which occurs several atomic units away from the molecule and requires large angular momenta, additional (aug-cc-pVXZ, X=T,Q) basis sets are strategically placed at `ghost' centres close to the regions of maximum positron density. 
To check convergence with respect to the number and location of these ghost centres, for each molecule we performed calculations including TZ or QZ bases on a successively increasing number of ghosts centres in different arrangements until the increase in binding energy fell below a few percent.
We found that including ghosts can increase binding energies by $\sim10\%$ in the polar molecules, and easily by $\sim30\%$ for the nonpolar ones, e.g., for CS$_2$ we obtained $\varepsilon_b=39$\,meV at $GW$@BSE+$\Gamma+\Lambda$ level with no ghosts, rising to $\varepsilon_b=68$\, meV with 16 additional ghosts. The use of higher angular momenta and more ghosts would further increase the binding energies. 
We also investigated the difference of using aug-cc-pVXZ for X=T,Q in the atomic centred and ghost bases, higher angular momenta in the even tempered basis. Some improvement was noted moving from X=T to Q, and from including $g$ states in addition to $f$,  to a level of a 5-10\% in polar molecules, and 10-30\% in nonpolars. Overall, good convergence with respect to both the electron and positron bases was observed (see e.g., Extended Data Fig.~2). 

The coefficients $C$ in the expansion of the positron wavefunction in Gaussians are found by solving the Roothaan equations $\bm{F}^{\pm}\bm C^{\pm}=\bm S^{\pm} \bm C^{\pm}\bm\varepsilon^{\pm}$, where $F^{\pm}$ is the Fock matrix and $S$ is the overlap matrix.
The one-body and two-body Coulomb integrals of the Fock matrix are calculated using the McMurchie Davidson algorithm\cite{mcmurchie}.
We eliminate linearly-dependent states by excluding eigenvalues $<10^{-5}$ of the overlap matrices (typically $\lesssim5\%$ of the states). 
In practice, to minimise the basis dimensions we transform all quantities to a spherical harmonic Gaussian basis (for a given angular momentum, the number of Cartesian Gaussians is greater than or equal to the number of spherical Harmonic Gaussians)\cite{Schlegel:1995}. 
Solution of the Roothaan equations yield bases of electron and positron Hartree-Fock MOs $\{\varphi_{\alpha}^{\pm}(\bm{r})\}$ (which include ground and other negative energy states, and discretized continuum states)  with which the self-energy diagrams can be constructed (see below for details). 

Expanding the positron Dyson wavefunction (see Eqn.~1 of main text) in the positron HF MO basis as $\psi_{\varepsilon}({\bf r})= \sum_{\nu} D^{\varepsilon}_{\nu} \varphi^+_{\nu}({\bf r})$ transforms the Dyson equation to the linear matrix equation $\bm{HD}=\bm{\varepsilon D}$, where $ \langle \nu_1 | H | \nu_2\rangle = \varepsilon_{\nu_1} \delta_{\nu_1\nu_2} + \langle \nu_1 |\Sigma_{\varepsilon} | {\nu_2}\rangle$. 
Note that we calculate the full self-energy matrix including off-diagonal terms. Such a non-perturbative approach is essential for nonpolar molecules, where binding is enabled exclusively by correlations. 
In practice, to obtain the self-consistent solution to the Dyson equation, we calculate the self energy at a number of distinct energies $E_i$ spanning the true binding energy $\varepsilon_b$, with the latter determined from the intersection of the $\varepsilon_b(E_i)$ data with the line $\varepsilon_b(E)=E$. 

\subsubsection*{Constructing the positron-molecule self energy via solution of the BSE equations.}
As discussed in the main text (Fig.~1), we consider 
three contributions to the irreducible self energy of the positron in the field of the molecule: 
 $\Sigma^{GW}$ (which describes polarization, screening and electron-hole interactions); $\Sigma^{\Gamma}$ (which describes the non-perturbative process of virtual-positronium formation); and $\Sigma^{\Lambda}$ (which includes the infinite ladder series of positron-hole interactions). 
In practice, we construct the individual contributions by first solving the respective Bethe-Salpeter equations (see Extended Data Fig.~1) for the electron-hole polarization propagator $\Pi$, the two-particle positron-electron propagator $G^{\rm ep}_{\rm II}$ and the positron-hole two-`particle' propagator $G_{\rm II}^{\rm ph}$\cite{mbtexposed}. Their general form is
${\bm L}({\omega}) = {\bm L}^{(0)}(\omega) + {\bm L}^{(0)}(\omega)\bm{KL}(\omega)$ 
where the ${\bm L}^{(0)}$ are non-interacting two-body propagators 
and $K$ are the interaction kernels\cite{FetterWalecka,mbtexposed,pqc} [e.g., see Extended Data Fig.~1\,e for the BSE for the electron-hole polarization propagator $\Pi$]. 
In the excitation space of pair product HF orbitals ${\bm L} = 
\left(\bm{C}\omega-\bm{H}\right)^{-1} =
{\bm \xi} (\omega-{\bm \Omega})^{-1}{\bm \xi}^{-1}{\bm C}^{-1}$, where the pair transition amplitudes 
$\xi$ are the solutions of the pseudo-Hermitian linear-response generalised eigenvalue equations  \cite{nucmbt,pqc,HeadGordon:2005}
$\bm H {\bm \xi} = {\bm C}{\bm \xi}{\bm \Omega}$,  ${\bm \xi}^{\dag}{\bm C}{\bm \xi}={\bm C}$,
where 
%\begin{linenomath}
\begin{equation}
{\bm H} =
\left(\begin{array}{cc}
{\bm A} & {\bm B} \\
{\bm B}^{\ast} & {\bm A}^{\ast}
\end{array}\right)~~;~~
{\bm \xi}=
\left(\begin{array}{cc}
{\bm X} & {\bm Y}^{\ast} \\
{\bm Y} & {\bm X}^{\ast}
\end{array}\right)~~;~~
{\bm C}=
\left(\begin{array}{cc}
{\bm 1} & {\bm 1} \\
{\bm 0} & {\bm -1}
\end{array}\right)~~;~~
{\bf \Omega} = \left(\begin{array}{cc}
{\bm\Omega_+} & 0 \\
0 & {\bm \Omega_-}
\end{array}\right),
\end{equation}
%\end{linenomath}
for excitation energies $\Omega^{\alpha}_+$ and $\Omega^{\alpha}_-$, which are labelled by $\alpha=1,\dots, \dim(\bf A)$. 
Here the ${\bm A}$ and ${\bm B}$ matrices depend on the particular two-particle propagator ${\bm L}$ under consideration and the approximation used for it, (see Extended Table 4 for the explicit matrix elements): note that ${\bm B= \bm 0}$ for the two-particle propagators involving the positron, since the vacuum state for the diagrammatic expansion is that of the $N$-electron molecule, and thus there are no positron holes and only time-forward positron propagators.
To determine the amplitudes, we employ the parallel diagonalisation algorithm of Shao \cite{Shao:2016}, which exploits a similarity transform that gives the eigenvalues of ${\bm C}^{-1}{\bm H}$ as the square roots of the eigenvalues of $({\bm A}+{\bm B})({\bm A}-{\bm B})$ (thus requiring matrices of dimension of the $\bm{A}$ block, i.e., half of the full BSE matrix dimension)  
to obtain ${\bm X} = \frac{1}{2}\left( {\bm L}_2{\bm U} +{\bm L}_1\bm{V}\right){\bm \Omega}_+^{-1/2}$ and ${\bm Y} = \frac{1}{2}\left( {\bm L}_2{\bm U} -{\bm L}_1\bm{V}\right){\bm\Omega}_+^{-1/2}$, via the Cholesky decompositions ${\bm A}+{\bm B} = {\bm L}_1{\bm L}_1^T$ and ${\bm A}-{\bm B} = {\bm L}_2{\bm L}_2^T$, and the singular value decomposition ${\bm L}_2{\bm L}_1^T={\bm U}\bm{\Omega}{\bm V}^T$.
The positron-molecule self-energy matrix elements can then be written as 
%\begin{linenomath}
\begin{align}
\langle \nu_1|\Sigma_E^{GW}|\nu_2\rangle&=\sum_{\alpha,\nu_3}\frac{w_{\nu_1 \nu_3}^{\Pi,\alpha} w_{\nu_2\nu_3}^{\Pi,\alpha}}{E-\varepsilon_{\nu_3}-\Omega^{\Pi}_{+,\alpha}+{\rm i}\eta},\\
\langle \nu_1|\Sigma_E^{\Gamma}|\nu_2\rangle&=\sum_{\alpha,n}\frac{{w}_{\nu_1 n}^{\Gamma,\alpha}{w}_{\nu_2n}^{\Gamma,\alpha}}{E-\Omega_\alpha^{\Gamma}+\varepsilon_{n}+{\rm i}\eta}-\langle \nu_1|\Sigma_E^{(2)}|\nu_2\rangle,\\
\langle \nu_1|\Sigma_E^{\Lambda}|\nu_2\rangle&=\sum_{\alpha,\mu}\frac{w_{\nu_1 \mu}^{\Lambda,\alpha}w_{\nu_2\mu}^{\Lambda,\alpha}}{E-\Omega_\alpha^{\Lambda}-\varepsilon_{\mu}+{\rm i}\eta}-\langle \nu_1|\Sigma_E^{(2)}|\nu_2\rangle,
\end{align}
%\end{linenomath}
where $\nu_1$, $\nu_2$ and $\nu_3$ denote positron indices and $\mu$ and $n$ denote electron excited states and holes respectively, and ${\Sigma}^{(2)}$  -- which results from the $\Pi^{(0)}$ contribution to $\Sigma^{GW}$  and is present in both $G^{\rm ep}_{\rm II}$ and $G_{\rm II}^{\rm ph}$ --
is subtracted to prevent double counting, 
and
%\begin{linenomath}
\begin{align}\label{eqn:ws}
w_{\nu_1 \nu_3}^{\Pi,\alpha}=\sum_{\mu n}(\nu_1 \nu_3|\mu n)(X^{\Pi,\alpha}_{\mu n}+Y^{\Pi,\alpha}_{\mu n})~~~;~~~
w_{\nu_1 n}^{\Gamma,\alpha}=\sum_{\mu\nu_3}(\nu_1 n|\nu_3 \mu)X^{\Gamma,\alpha}_{\nu_3\mu}~~~;~~~
w_{\nu_1 \mu}^{\Lambda,\alpha}=\sum_{n \nu_3}(\nu_1 \mu|\nu_3 n)X^{\Lambda,\alpha}_{\nu_3n}.
\end{align}
%\end{linenomath}
The total self energy is thus calculated as $\Sigma=\Sigma^{GW}+\Sigma^{\Gamma}+\Sigma^{\Lambda}$. % with experiment and previous theory for each molecule in turn.
%\begin{comment}
%To understand the accuracy of our approach, we first note that 
Such addition of the individual channels is routine in atomic many-body theory calculations \cite{boylepindzola,Gribakin:2004,DGG_posnobles} and in condensed matter, e.g., the fluctuation-exchange (`FLEX') approximation \cite{Reiningbook,Bickers:1989a,Bickers:1989b}. 
%In reality, however, the true self energy involves coupling of these channels. 
%For example, corrections to the virtual-positronium diagram in Fig.~\ref{fig:diags_jd} (b) may include insertions of the polarization propagator $\Pi$ between the propagating electron and hole, i.e., combining Fig.~\ref{fig:diags_jd} (a) and (b). Such coupling of self-energy terms can be taken account of approximately via the `parquet'  \cite{Reiningbook} or Fadeev\cite{Fadeevmol} equations, or algebraic diagrammatic construction ADC(3) method \cite{ADC3}, but their computational expense is beyond our current available resources. 
%Thus, although the ${GW@{\rm BSE}}+{\tilde{\Gamma}+\tilde{\Lambda}}$ approximation is technically our most sophisticated level of theory, the neglect of the couplings means that our true binding energy may be expected to be somewhere between the $GW@{\rm BSE}+{\Gamma+\Lambda}$ and $GW@{\rm BSE}+{\tilde{\Gamma}+\tilde{\Lambda}}$ results, with the latter representing the lower bound.
%\end{comment}

We implement the above in the massively-parallelised {\tt EXCITON+} code developed by us, heavily adapting the {\tt EXCITON} code \cite{exciton,patterson:2019,patterson:2020} to include  positrons and the many-body theory capability (calculation of the self energy and solution of the Dyson equation).
We employ density fitting \cite{dfCoul1,dfCoul2,dfCoul3,dfOlap1,dfOlap2,patterson:2020} (of the electronic density) to calculate the Coulomb integrals in the matrix elements of ${\bm A}$ and ${\bm B}$, in $w^{\Pi}$, $w^{\Gamma}$ and $w^{\Lambda}$, and positron-electron contact density, via a parallel implementation that assigns matrix elements involving auxiliary basis functions on distinct atomic centres to distinct processors, similar to that used in the {\tt MolGW} program \cite{MolGW_CPC}. 
The employment of density fitting reduces four-centre Coulomb integrals to products of three-centre Coulomb integrals and matrix elements of the Coulomb operator between atomic orbital basis functions. Thus, the memory scaling is $\sim N_-^2 M_-$, where $N_-$ is the total number of electron basis functions, and $M_-\gtrsim 3N_-$ is the number of electron auxiliary basis functions. The most computationally demanding part of our approach is in the calculation of the virtual-Ps self-energy contribution $\Sigma^{\Gamma}$. 
For this, $\dim{\bm A}=\dim X^{\Gamma} = N_{\nu}\times N_{\mu}$, the product of total number of positron MOs and excited electron MOs.  For the calculations considered here, $N_{\nu}$ ranged from 400--500 and $N_{\mu}$ from 300--400, resulting in $\dim X^{\Gamma} = 120,000$--200,000, and thus diagonalising the matrix of $(\dim X^{\Gamma})^2$ elements demanded between $\sim$100 GB and 1.5 TB of RAM. 
The calculations were performed on two AMD EPYC 128\,CPU\,@\,2\,GHz, 768GB RAM nodes of the United Kingdom Tier-2 supercomputer `Kelvin-2' at Queen's University Belfast. 
In contrast, the $GW$ calculations involve $\dim{\bm A}=\dim X^{\Pi} \le N_{\nu}\times N_{n}$, i.e., a maximum equal to the product of the number of occupied and excited electron MOs: in practice not all occupied orbitals need to be included because the tightly bound LOMOs are less susceptible to perturbation by the positron and have negligible contribution to the self energy. 
Thus, since $N_{n}\ll N_{\mu}<N_{\nu}$, \emph{ab initio} $GW${\rm @RPA/TDHF/BSE} calculations are considerably less  computationally expensive, and can be performed for molecules or clusters with $\sim$ 100 atoms, providing at least lower bounds on the positron binding energies. 
Moreover, as discussed  in the main text (see Fig.~3\,c of main text and Extended Data Fig.~3) and demonstrated for nucelobases (Table 2 of main text), the virtual-Ps formation contribution can be approximated by scaling the $\Sigma^{(2)}$ self energy by the strength parameter ratio $g\equiv\mathcal{S}^{(2+\Gamma)}/\mathcal{S}^{(2)}$, viz.~$\Sigma\approx g\Sigma^{(2)}+\Sigma^{\Lambda}$, thus enabling computationally relatively inexpensive binding-energy calculations that account for virtual-Ps formation for molecules of $\sim$ 100 atoms.
\emph{Ab initio} calculations for larger molecules including the virtual-positronium self energy will be feasible with additional computational resources, as would calculations using different truncated product spaces of excited electron and positron MOs and extrapolating to the basis set limit.

%\subsubsection*{Strength of .}

\subsubsection*{Improving the accuracy of calculations}
As mentioned in the previous section, the computationally intensive calculations presented here were performed using relatively modest computational resources. Access to national supercomputing facilities would enable more complete basis sets and further exploration of the effect of ghost basis centres.
Numerical accuracy can also be systematically improved in a number of ways. 
Exploiting the molecular point group symmetry via symmetry adapted bases and employing integral screening techniques would improve the efficiency of the calculations, enabling more complete basis sets to be used. This would ultimately improve the description of the correlations (particularly in generating higher angular momenta for improved description of the virtual-positronium formation process).
The calculation of the positron-molecule self energy can be improved by implementing a self-consistent diagram approach in which the positron-molecule self energy is built from $GW$ calculated electron and positron Dyson orbitals rather than HF ones \cite{mbtexposed,qsgw}, and/or by coupling the three self-energy channels $\Sigma^{GW}$, $\Sigma^{\Gamma}$ and $\Sigma^{\Lambda}$ by approximating the three-particle propagators via the Fadeev\cite{Fadeevmol}, parquet\cite{Reiningbook} or ADC(3) methods \cite{ADC3} (expected to be computationally feasible for small molecules using national supercomputing facilities). 
Moreover, the diagrammatic series should be amenable to a diagrammatic Monte Carlo\cite{diagmcbook,bolddiagMC} prescription, a stochastic simulation method that enables the effective summation of many more (classes of) diagrams than considered here.

\subsubsection*{Positron annihilation rate in the bound state}

\noindent 
%In addition to the positron-molecule binding energies, the self-consistent 
Solution of the Dyson equation also yields the positron-bound state wavefunction $\psi_{\varepsilon}$. Using it,
%\end{eqnarray}
the 2$\gamma$ annihilation rate in the bound state $\Gamma= \pi r_0^2c\delta_{ep}$ ($\Gamma[{\rm ns}^{-1}]= 50.47\,\delta_{ep}[{\rm a.u.}])$, whose inverse is the lifetime of the positron-molecule complex with respect to annihilation, can be calculated. 
%the 2$\gamma$ annihilation rate in the bound state
Here $r_0$ is the classical electron radius, $c$ is the speed of light and $\delta_{ep}$ is the electron-positron \emph{contact density}
%\begin{equation}
%\delta_{ep}=\sum_{i=1}^{N_e} \int d {\bm r} d {\bm r}_1 \dots  d {\bm r}_{N_e} \delta({\bm r}-{\bm r}_i) |\Psi({\bm r}_1\dotsi {\bm r}_{N_e},{\bm r})|^2,
%\end{equation}
%also known as the contact density, and $\Psi({\bm r}_1\dotsi {\bm r}_{N_e},{\bm r})$ is the total wavefunction of the positron-molecule bound state, normalised to unity.
\begin{equation}\label{eqn:delta_ep}
\delta_{ep}=\sum_{n=1}^{N_e}  \gamma_{n} \int   |\varphi_n({\bm r})|^2 |\psi_{\varepsilon}({\bm r})|^2d {\bm r}.
\end{equation}
Here the sum is over all occupied electron MOs with wavefunctions $\varphi_n$, and $\gamma_n$ are MO dependent enhancement factors that account for the short-range electron-positron attraction \cite{DGG:2015:core,DGG:2017:ef}. 
%They quantify the effect of corrections to the annihilation vertex in the Dyson expansion of the annihilation amplitude, and 
Recent many-body calculations for atoms by one of us determined them to follow a physically motivated scaling with the ionization energy \cite{DGG:2015:core,DGG:2017:ef}  
$\gamma_n = 1+ \sqrt{ {1.31}/{|\varepsilon_n|}} + \left({0.834}/{|\varepsilon_n|} \right)^{2.15}$, which we assume to hold here. 
%In reality, 
The positron Dyson wavefunction is a quasiparticle wavefunction that is the overlap of the wavefunction of the $N$-electron ground state molecule with the fully-correlated wavefunction of the positron plus $N$-electron molecule system\cite{mbtexposed}. It is normalised as 
\begin{eqnarray}\label{eqn:aval}
\int |\psi_{\varepsilon}({\bm r})|^2 d{\bf r}= \left(1-{\partial \varepsilon}/{\partial E}\right|_{\varepsilon_b})
^{-1} \equiv a <1,
\end{eqnarray} 
which estimates the contribution of the `positron plus molecule in the ground state' component to the positron-molecule bound state wavefunction, i.e., the degree to which the positron-molecule bound state is a single-particle state, with smaller values of $a$ signifying a more strongly-correlated state. Fig.~4 and Extended Data Figs.~4 present contact density data. Extended Data Fig.~4\,a shows the individual MO contribution to the contact density as a function of the MO ionisation energy: similar to Fig.~3 of the main text (contribution of strength parameters from individual MOs), overall the contact density increases as the ionisation energy decreases: the positron overlap is greater with the more diffuse electronic HOMOs. However, MOs below the HOMO can in fact dominate, e.g., acetonitrile, as shown further in Fig~4\,a, b and c. 
%Extended Data Fig.~4\,b shows that the total contact densities are to a good approximation proportional to the square root of the binding energy (this results from the normalisation of the wavefunction). 
%including contributions from individual molecular orbitals, and as a function of ionization energy. The renormalization constants are presented in Extended Data Table 2. %and $a$ values for all molecules considered in this work. 
}%end small methods

\section*{Data availability}
All relevant data generated and analysed during this work are available from DGG on reasonable request. (On acceptance, they will be made freely available on the Queen’s University Belfast official data repository https://pure.qub.ac.uk/en/datasets/).

\section*{Code availability}
The results presented in this study were generated using the program {\tt EXCITON+} that was newly developed by the authors, heavily adapting the {\tt EXCITON} code to incorporate positrons and many-body theory.  The source code can be made freely available by DGG on reasonable request. We intend to detail the code in a subsequent article.

%\newpage
%\vspace*{45ex}
%\renewcommand\refname{Methods References}
%\bibliographystyle{naturemag}
%\bibliography{posmol-mbt-binding.bib}

%\renewcommand\refname{Methods References}
%\bibliographystyle{naturemag}
%\bibliography{dgreen_all.bib}

%\newpage
\section*{Acknowledgements}
We thank James Danielson and Cliff Surko (University of California San Diego) for providing unpublished data for formamide, and Alin Alena and Martin Plummer (STFC Scientific Computing Laboratory, Daresbury United Kingdom), Ian Stewart and Luis Fernadez Menchero (Queen's University Belfast) for high-performance computing support. D.G.G.~additionally thanks Gleb Gribakin for many insightful discussions and for drawing his attention to this problem, and Andrew Swann for comments on the manuscript. 
This work was supported by D.G.G.'s 
European Research Council grant 804383 ``ANTI-ATOM". 

\section*{Author Contributions}
B.C.~and D.G.G. led, supported by J. H., C.M.R. and C.H.P, the development of the positron-molecule many-body theory and its computational implementation in the newly-developed {\tt EXCITON+} code, which was based on the {\tt EXCITON} code developed by C.H.P.~for \emph{electronic} structure calculations. % but heavily adapted in this work by B.C., J.H., C.M.R, and D.G.G to incorporate positrons and the positron many-body theory (calculation of electron and positron self energies and solution of Dyson equations).
J.H.~performed the majority of calculations, supported by C.M.R.\, 
D.G.G.~led the analysis, supported by J.H.
%All authors contributed to the manuscript preparation.
D.G.G.~additionally conceived and supervised the work and drafted the manuscript, which all authors contributed to editing.\\

\noindent The authors declare no competing interests.

\section*{Additional information}
Correspondence should be addressed to D.G.G. (d.green@qub.ac.uk).

\newpage
\section*{Extended data}

\setcounter{figure}{0}  
\renewcommand{\figurename}{\small {\bf Extended Data Figure}}
\begin{figure}[hp!!]
\centering
\includegraphics*[width=0.9\textwidth]{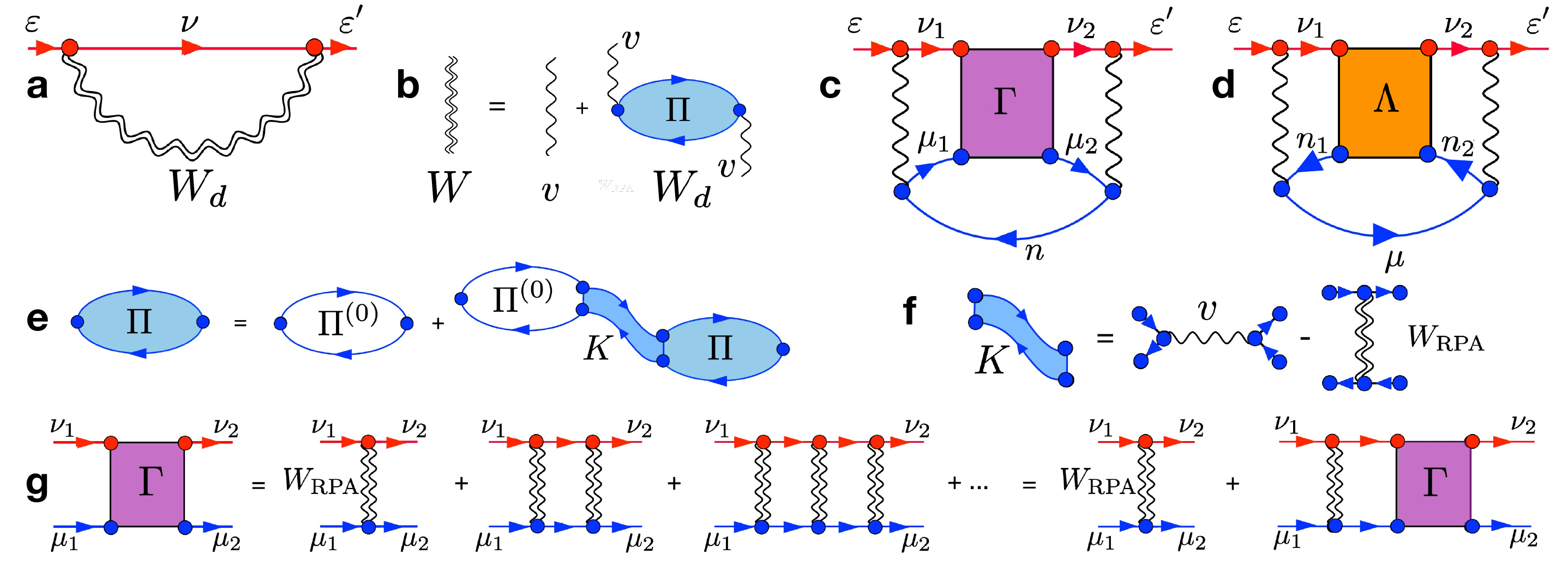}~~~
\caption*{\footnotesize{\bf Extended Data Figure 1. The main contributions to the positron-molecule self energy, including the two-particle propagators.} 
{\bf a}, the $GW$ diagram involves the positron Green's function $G_{\nu}$ and the dynamic part (due to the absence of an electron-positron exchange interaction) of the screened Coulomb interaction $\boldsymbol{W}_{\rm d}=\boldsymbol{ v\Pi v}$ (bold denotes operator form), where $\Pi$ is the electron-hole polarization propagator (see {\bf b}). 
It satisfies the Bethe-Salpeter equation (diagram {\bf e}) with kernel $K=v-W_{\rm RPA}$ (diagram {\bf f}), where $W_{\rm RPA}= v+ W_{\rm d, RPA}$ is the screened electron-hole Coulomb interaction calculated in the random phase approximation.
% and $v$ and $v_{\rm exch.}$ are direct and exchange Coulomb interactions. 
Setting $K=0$ results in the bare polarisation entering $W$ only, and gives the $\Sigma^{(2)}$ approximation, so-called as it is a second-order diagram in the electron-positron Coulomb interaction. Setting $K=v$, the direct part of the Coulomb interaction only, gives the `random phase approximation' ($GW$@RPA). Setting $K=v-v_{\rm exch}$, i.e., including exchange which gives rise to interactions within the bubbles, yields the `time-dependent Hartree-Fock' approximation ($GW$@TDHF). Using screened Coulomb interactions in the exchange term is `Bethe-Salpeter' approximation ($GW$@BSE). %with $W_d$ calculated using the polarization propagator kernel $K=v$, 
%{\bf (c)} The virtual-positronium contribution  $\Sigma^{\Gamma}$, which includes the summed infinite ladder-diagram series of screened electron-positron interactions,  the `${\Gamma}$ block' [diagram (g)] and (d) the positron-hole ladder series contribution. 
%See also Extended Data Fig.~1.}
%Depending on the 
{\bf g}, the summed infinite ladder series of screened electron-positron interactions (`$\Gamma$-block'). The $\Lambda$-block (diagram {\bf d}) is the ladder series of positron-hole interactions, and it satisfies a linear integral equation of the same form as that shown in {\bf g}.
\label{fig:acetocon}}
\end{figure}

%\newpage

\renewcommand{\tablename}{\footnotesize{Extended Data Table}}

\begin{table*}[hp!!]
\footnotesize
\hrule
\smallskip
\smallskip
\noindent{~\bf ~\small Extended Data Table 1: Calculated polarizabilities (in \AA$^3$) and ionization energies (in eV).}\\[-1ex]
\hrule
\smallskip
\begin{tabular}{lccccc@{\hskip 10pt}ccc@{\hskip 10pt}ccc}
&\multicolumn{5}{c}{Isotropic polarizability $\alpha$ (\AA$^3$) }&\multicolumn{3}{c}{BSE Pol. $\alpha$ (\AA$^3$)}&\multicolumn{3}{c}{ionization energy (eV)}\\
\cmidrule(lr){2-6}\cmidrule(lr){7-9}\cmidrule(lr){10-12}
& HF ($\Pi^0)$ & RPA & TDHF & BSE & Ref.$^{\blue{\dag}}$ & $xx$&$yy$&$zz$&HF & $GW$@RPA & Ref.$^{\blue{\dag}}$\\
\hline
{\bf Polar molecules}\\
LiH &2.19	& 1.72	&3.61&3.50 & 3.68 & 1.83&1.83&1.60&8.19 & 8.25 & 7.70\\
%HCN &2.26&1.50&2.40	&2.37 & 2.46 & 13.75& 14.02 & 13.60\\
Acetonitrile &3.93&	2.66&	4.29&	4.24 & 4.40 &1.77&1.77&2.83& 12.46 & 12.58&12.20\\
Propionitrile &5.38	&3.69	&5.97&	5.90 & 6.24 &2.54&2.65&3.66& 12.37&12.41 & 11.84\\
Acetone &5.04&	3.55&	5.74	&5.74 & 6.33 &2.38&3.07&3.15& 11.18&10.17 & 9.70\\
Propanal&5.03	&3.54	&5.71&	5.70 & 6.50 &2.94&2.52&3.08& 11.39&10.44 & 9.96\\
Acetaldehyde&3.57	&2.53	&4.10	&4.12 & 4.59 &1.71&2.06&2.41& 11.53&10.55 & 10.23\\
Formamide&3.17&	2.25	&3.61&	3.68 & 4.08 &1.39&1.92&2.21& 11.58&11.02 & 10.16\\[0.5ex]
{\bf Nonpolars}\\
CS$_2$&8.74	&4.96&	8.14	&8.06 & 8.74 &2.70&2.70&6.70& 10.13&10.46 & 10.07\\
CSe$_2$&11.84&6.53&10.85	&10.70 & - &3.44&3.44&9.18& 9.33 &9.70 & -\\
Benzene&9.79	&6.22	&9.88	&9.85 & 10.00 &5.74&5.74&3.31& 9.21&9.52 & 9.24\\[0.1ex]
\hline
\end{tabular}
\caption*{\footnotesize $^{\blue{\dag}}$Reference values from Methods Ref.~[\citen{CRC97}]. Molecules are oriented such that the main axis of symmetry, or the main bond (C-O, C-N), above which the positron density is localised, is along $z$.  The isotropic value is given by a sum of $xx$, $yy$, and $zz$ terms multiplied by $2/3$. Note that the $zz$ components have larger differences between molecules than isotropic polarizabilities, e.g., for propionitrile, acetone and propanal the isotropic polarisabilities are within ~1\% of each other, whereas the $zz$ components differ by $\sim 15\%$.
Ionisation energies calculated at the $GW$@RPA level were performed using the diagonal approximation for the electron-molecule self energy  $\Sigma^{(-)}$, i.e.,
$\tilde{\varepsilon}_{\mu}=\varepsilon_{\mu}+
%\langle\mu'|\Sigma^{(-)}_E|\mu\rangle\approx 
Z\langle\mu|\Sigma^{(-)}_{\varepsilon_{\mu}}|\mu\rangle$, where $Z\equiv (1-\partial\Sigma_{E}/\partial E)^{-1}|_{\varepsilon_{\mu}}$.\cite{Reiningbook}}
%\end{center}
\label{default}
\end{table*}%

\renewcommand{\tablename}{}
\begin{table*}[pt!]
\footnotesize
\hrule
\vspace*{5pt}
\caption*{{\bf  \small Extended Data Table 2: Positron binding energies in the $GW$ approximation (meV), dimensionless correlation-potential strength parameters and Dyson wavefunction renormalisation constants $a$}\label{table:be}}
\vspace*{-4ex}
\begin{center}
\begin{tabular}{l@{\hskip10pt} c@{\hskip5pt} c@{\hskip5pt} c@{\hskip5pt} c@{\hskip5pt}c@{\hskip25pt} c@{\hskip5pt} c@{\hskip6pt} c@{\hskip6pt}c@{\hskip6pt}c@{\hskip16pt}c@{\hskip26pt}c@{\hskip0pt}}
\hline\\[-1.5ex]
%&\multicolumn{4}{c}{$GW$} &\multicolumn{3}{c}{$GW$@BSE+}  & & \\
&&\multicolumn{4}{c}{$\varepsilon_b$ calculated using $\Sigma^{GW@}$} &\multicolumn{5}{c}{Strength parameter} & ~~~Renorm.$a$\\
\cmidrule(lr){3-6}\cmidrule(ll){7-11}\cmidrule(ll){12-12}
			& HF		& $\Sigma^{(2)}$ 	& RPA 	& TDHF 	& BSE 	& 	 $\mathcal{S}^2$  &	 $\mathcal{S}^{GW}$& $\mathcal{S}^{\Gamma}$ & $\mathcal{S}^{\Lambda}$ &$\mathcal{S}^{GW+\Gamma+\Lambda}$ & $~~~\Sigma^{GW+\Gamma+\Lambda}$ \\
\hline
{\bf Polars} 	&		&			&				&					&										\\
LiH  			&130 	& 434 		&336 	& 542 		& 518 	& 4.4 & 4.9 & 3.8& $-$0.9 &	 7.8 &	0.860	\\
%HCN		&16.0&14.0& 2		&27			&7		&24		&23		&111			& 67					&62								& - 		&2		&35		& 38		& 8	\\%	&31-82\\
Acetonitrile	&15		&120			& 59		& 112	& 109	&8.3 &7.5&2.8&$-$1.2 &9.1 &	 0.972	\\
Propionitrile	& 16		& 140		& 69		& 133	& 129	&10.3 &9.5&3.4&$-$1.3 &11.6 &	 0.968	\\
Acetone		& 3		& 67			&25		& 71		& 69	 &12.9&12.5&4.2&$-$1.7 &15.0 &	 0.977	\\
Propanal		& 1		& 44			&12		& 46		& 45	&11.1 &10.6&3.6&$-$1.6 &12.7 &	 0.980	\\
Acetaldehyde	& 2		& 35			&11		& 39		& 38	&9.3 &9.0&3.1&$-$1.3 &10.8 &	 0.984	\\
Formamide	& 12		&105			&58		& 109		& 108	&7.1 &6.7&2.3&$-$1.0 &8.0 &	 0.977	\\
{\bf Nonpolars}			&			&					&		& & & &  & & 	\\
CS$_2$		& $<0$	&$<0$		&$<0$	& $<0$	&$<0$	&11.5 &9.6&4.7&$-$1.8 &12.5 &	 0.959	\\
CSe$_2$		& $<0$	&9 			&$<0$	& $<0$	&$<0$	&11.9 &10.0&5.2&$-$1.7 &13.5 &	 0.931	\\
Benzene		&$<0$	&11			&$<0$	&	2	&	2	&15.0 &13.0&5.0&$-$2.1 &15.9 &	 0.951		\\
\hline\\[-5ex]
\end{tabular}
\end{center}
\renewcommand{\tablename}{}
\caption*{\footnotesize  Positron binding energies (in meV, complementary data to Table 1 of main text) calculated at the HF and various levels of the $GW$ approximation (see Extended Data Fig.~1): 
%Different approximations to $W$ arise from different approximations to the kernel $K$ of the electron-hole polarization propagator $\Pi$.
% compared with experiment and previous theory. (Also included are the calculated 
%isotropic polarisabilities $\alpha$ and ionisation energies $I$ calculated at the $GW$@BSE level of the theory: see Extended Table 2 for more details and comparison with reference values.) 
%Calculations are presented at the Hartree-Fock approximation, various levels of the $GW$ approximation (see Fig.~1): 
$\Sigma^{(2)}$ (bare polarisation propagator); 
RPA (random phase approximation); 
TDHF (time-dependent Hartree Fock approximation); and 
BSE (Bethe-Salpeter equation).
Dimensionless strength parameter of the correlation potential (defined in the main text) in different approximations to the positron-molecule self energy (see Fig.~3 of main text). Positive (negative) strength parameters denote attractive (repulsive) positron-molecule interactions. 
The final column gives the calculated positron Dyson wavefunction renormalization constants $a$ for the $\Sigma^{GW+\Gamma+\Lambda}$ calculation (see `Methods' Eqn.~7).
%and for $GW@$BSE plus the virtual-positronium formation ladder series ($\Sigma^{GW+\Gamma}$), 
%additionally the positron-hole ladder series ($\Sigma^{GW+\Gamma+\Lambda}$), and finally $\Sigma^{GW+\tilde{\Gamma}+\tilde{\Lambda}}$ which uses dressed Coulomb interactions in the ladder series.\\[2pt]
%\blue{$^\dag$}Experimental results are resonant annihilation measurements by Danielson, Surko and co-workers at UCSD\cite{Danielson10,Danielson12},  \blue{$^{\ast}$}except for formamide, for which the measurement is preliminary and a final value has yet to be determined \cite{pJames}.
%
%$^1$ H. A. Kurtz and K.  D. Jordan, J. Chem. Phys. 75, 1876 (1981),\\
%$^2$ K.  Strasburger, Chem. Phys.  Lett. 253, 49-52,(1996), \\
%$^3$ S. Bubin and L. Adamowicz,  J.  Chem.  Phys. 120,6051 (2004),\\
%$^4$ M. Tachikawa, Y. Kita and R. J. Buenker, Phys. Chem. Chem. Phys. 13, 2701 (2011),\\
%$^5$ M. Tachikawa, J. Phys. Conf. Ser. 488, 012053 (2014),\\
%$^6$ J. Romero et.al., J. Chem.Phys. 141 (2014), \\
%$^7$ M. Tachikawa, Y. Kita and R. J. Buenker, New J. Phys. 14, 035004 (2012),\\
%$^8$ K. Koyanagi et al., Phys. Chem. Chem. Phys. 15, 16208 (2013)
} %and additionally including the positron hole ladder series￼}
%The suggested best results are the HF + GW@BSE + ￼.... 
%The error on the experimental values is ~?meV... (can we mention the systematic error? Show calculated polarisaibilities: why is our CS2 polarisability ~8 instead of 10... multiplying 61*10/8 gets us to 75
\vspace*{-5pt}\hrule
\label{default}
\end{table*}%

\renewcommand{\tablename}{}
\begin{table}[hp!]
\footnotesize
\hrule
\smallskip
\smallskip
\noindent{{\bf \small Extended Data Table 3 : Positron-molecule annihilation contact densities (a.u.)}}\\[-1ex]
\hrule
\smallskip
%\vspace*{-1ex}%\hrule
\begin{center}
\begin{tabular}{l c@{\hskip18pt} c@{\hskip18pt} c@{\hskip18pt}c@{\hskip18pt} c@{\hskip18pt} c@{\hskip22pt} c }
%& $\alpha $ & $I$ 
%&\multicolumn{2}{c}{}&&\multicolumn{4}{c}{$GW$} &\multicolumn{4}{c}{$GW$@BSE+}\\
%\hline
			& $\delta_{ep}^{(0)}$ 	& $\delta_{ep}^{GW}$ 	& $\delta_{ep}^{GW+\Gamma}	$ 	& $\delta_{ep}^{GW+\Gamma+\Lambda}$ 	& $\delta_{ep}^{GW+\tilde{\Gamma}+\tilde{\Lambda}}$ 	&	$\delta_{ep}^{GW+\tilde{\Gamma}+\tilde{\Lambda}\dag}$ \\ 	
\hline
{\bf Polars}	&		&			&		&		&		&			&			\\
LiH  			& 4.56[$-$4]&	 2.60[$-$2]	& 4.92[$-$2]	& 4.34[$-$2] &  4.14[$-$2] &4.15[$-$2]		\\
%HCN			& 		&			&		&		&		&			&					& 								&			\\
Acetonitrile	& 6.56[$-$5]&	5.67[$-$3]&1.62[$-$2]&1.14[$-$2]	&1.06[$-$2]	&1.11[$-$2]	\\
Propionitrile	& 7.38[$-$5]&	6.42[$-$3]	&1.68[$-$2]&	1.26[$-$2]&1.18[$-$2]&1.24[$-$2]	\\
Acetone		& 1.94[$-$5]&	5.75[$-$3]&	1.60[$-$2]&	1.10[$-$2]	&1.04[$-$2]&1.12[$-$2]\\
Propanal		& 9.14[$-$6]&	4.27[$-$3]&	1.36[$-$2]&	9.45[$-$3] &8.68[$-$3]&9.21[$-$3] \\
Acetaldehyde	& 1.23[$-$5]&	3.38[$-$3]&	1.12[$-$2]	&7.38[$-$3]&6.91[$-$3]&7.40[$-$3] \\
Formamide	& 8.64[$-$5]&	7.14[$-$3]	&1.65[$-$2]&	1.21[$-$2]	&1.18[$-$2]&1.23[$-$2]\\
{\bf Nonpolars}&	&	&			&					&								&	&		\\
CS$_2$	& -	&	-	&		1.22[$-$2]&	7.08[$-$3]&	5.62[$-$3]	&6.63[$-$3]	\\
CSe$_2$		&-	&	-	&		1.91[$-$2]&	1.02[$-$2]&	8.31[$-$3]	&9.59[$-$3]	\\
Benzene		&-&	1.21[$-$3]&	2.09[$-$2]	&1.35[$-$2]	&1.17[$-$2]	&1.32[$-$2]	\\[0.5ex]
\hline\\[-5ex]
\end{tabular}
\end{center}
\renewcommand{\tablename}{}
\caption*{\footnotesize Electron-positron contact densities calculated from Eqn.~\ref{eqn:delta_ep} including molecular-orbital-dependent enhancement factors and Dyson orbital renormalisation constants $a$ (Eqn.~\ref{eqn:aval}), in different approximations to the positron Dyson wavefunction: 
$\delta^{(0)}_{ep}$, using HF positron wavefunction; 
$\delta^{GW}_{ep}$, using the Dyson wavefunction calculated with the $GW$@BSE self energy; 
$\delta^{GW+\Gamma}_{ep}$, using the Dyson wavefunction calculated with the $GW$@BSE plus virtual-positronium  self energy; 
$\delta^{GW+\Gamma+\Lambda}_{ep}$, using the Dyson wavefunction calculated with the $GW$@BSE plus virtual-positronium  plus positron-hole self energy with unscreened Coulomb interactions in the $\Gamma$ and $\Lambda$ ladders;
$\delta^{GW+\tilde{\Gamma}+\tilde{\Lambda}}_{ep}$, using screened Coulomb interactions in the ladders and $^{\dag}$additionally using $GW$ energies in the diagrams. 
% accounting for the short-range electron-positron attraction \cite{DGG:2015:core,DGG:2017:ef}.  
Numbers in brackets indicate powers of 10. Hyphens denote approximations in which the positron does not bind. }
\vspace*{-4pt}%\hrule
\label{contact_densities}
\end{table}%

\begin{table}[p] 
\small
\hrule
\smallskip
\smallskip
\noindent{~\bf Extended Data Table 4: Matrix elements of the Bethe-Salpeter linear response Hamiltonian.}\\[-1ex]
%\smallskip
\hrule
\smallskip
\centering\begin{tabular}{lcll}
Method      				&   	$p_1p_2,p_3p_4$     & $ A_{(p_1p_2),(p_3p_4)}  $                                           	& $B_{(p_1p_2),(p_3p_4)}$          \\[2pt]
\hline\\[-2ex]
$GW$@HF       			&	$\mu n, \mu' m$ 	& $(\epsilon_{\mu} - \epsilon_n) \delta_{\mu\mu'} \delta_{nm}$           												& 0                    \\                              
$\phantom{GW}$@RPA      	& 	$\mu n, \mu' m$ 	& $(\epsilon_{\mu} - \epsilon_n) \delta_{\mu\mu'} \delta_{nm} + 2 (\mu n |m\mu')$     										& $2(\mu n|m\mu')$                 \\                              
$\phantom{GW}$@TDHF      	&	$\mu n, \mu' m$  	& $(\epsilon_{\mu} - \epsilon_n) \delta_{\mu\mu'} \delta_{nm} + 2 (\mu n |m\mu') - (\mu\mu'|mn) $ 							&  $2(\mu n|m\mu' )   - (n\mu' |m\mu)$ \\
$\phantom{GW}$@BSE       	&	$\mu n, \mu' m$  	& $(\widetilde{\epsilon_{\mu}} - \widetilde{\epsilon_n}) \delta_{\mu\mu'} \delta_{nm} + 2 (\mu n |m\mu') - (\mu\mu' |W| mn)$  		&  $2(\mu n|m\mu' )   - (n\mu'|W|m\mu)$\\                                                       
$\Sigma^{\Gamma}$ (virtual-Ps)   				&	$\nu\mu,\nu'\mu'$	& $(\widetilde{\epsilon_{\nu}} + \widetilde{\epsilon_{\mu}}) \delta_{\nu\nu'} \delta_{\mu\mu'} - (\nu\nu'|W|\mu\mu')$  				& 0\\
$\Sigma^{\Lambda}$ (positron-hole)  				&	$\nu n, \nu' m$		& $(\widetilde{\epsilon_{\nu}} - \widetilde{\epsilon_n}) +  (\nu \nu'|W|mn)$ 												& 0\\[0.5ex]
\hline\\[-4ex]
\end{tabular}
\caption*{\footnotesize Elements of the $\bm A$ and $\bm B$ blocks of the linear-response Hamiltonian matrices that result from the BSE equations (`Methods' Eqn.~1) for the electron-hole propagator, 
the positron-electron propagator, and the positron-hole propagator. Chemists’ notation for Coulomb matrix elements in the MO basis is used
$(\nu\mu|\nu'\mu')=\int{\rm d}{\bm r}{\rm d}{\bm r}'\varphi_{\nu}^{*}({\bm r})\varphi_{\mu}({\bm r})v({\bm r},{\bm r}')\varphi_{\mu'}^{*}({\bm r}')\varphi_{\nu}({\bm r}')$, where
($\mu$ and $\mu')$, ($n$ and $m$) and ($\nu$ and $\nu')$ denote electron particles, electron holes and positron particles respectively.
Factors of two arise from summation over spin, and tildes on energy eigenvalues for BSE denote that these are calculated at the level of $GW$@RPA.
For the virtual-Ps and positron-hole matrices, $\bm B=\bm0$ because there are no positron `holes'  in the $N$-electron ground-state molecule vacuum-state, and thus only time-forward diagrams are present in the positron single-particle propagator and two-particle propagators (i.e., here the `Tamm-Dancoff approximation' is exact for positrons). 
Matrix elements of the dressed Coulomb interaction $W=v+ W_d$ (Extended Data Fig.~1 b), where $W_d=v\Pi^{\rm RPA}v$ is the dynamic part determined from the polarization propagator in the random phase approximation, are determined as
$W_{\mu n, \mu' m} = (\mu n|\mu'm) + \sum_\alpha w_{\mu n}^\alpha w_{\mu' m}^\alpha \left[({\omega - \Omega_+^{\alpha} + i\eta})^{-1} - ({\omega + \Omega_+^{\alpha} - i\eta})^{-1} \right]$.\\\\
}
\label{tab1}
\end{table}

\renewcommand{\figurename}{\small {\bf Extended Data Figure}}
\begin{figure}[hp!!]
\centering
\includegraphics*[width=0.49\textwidth]{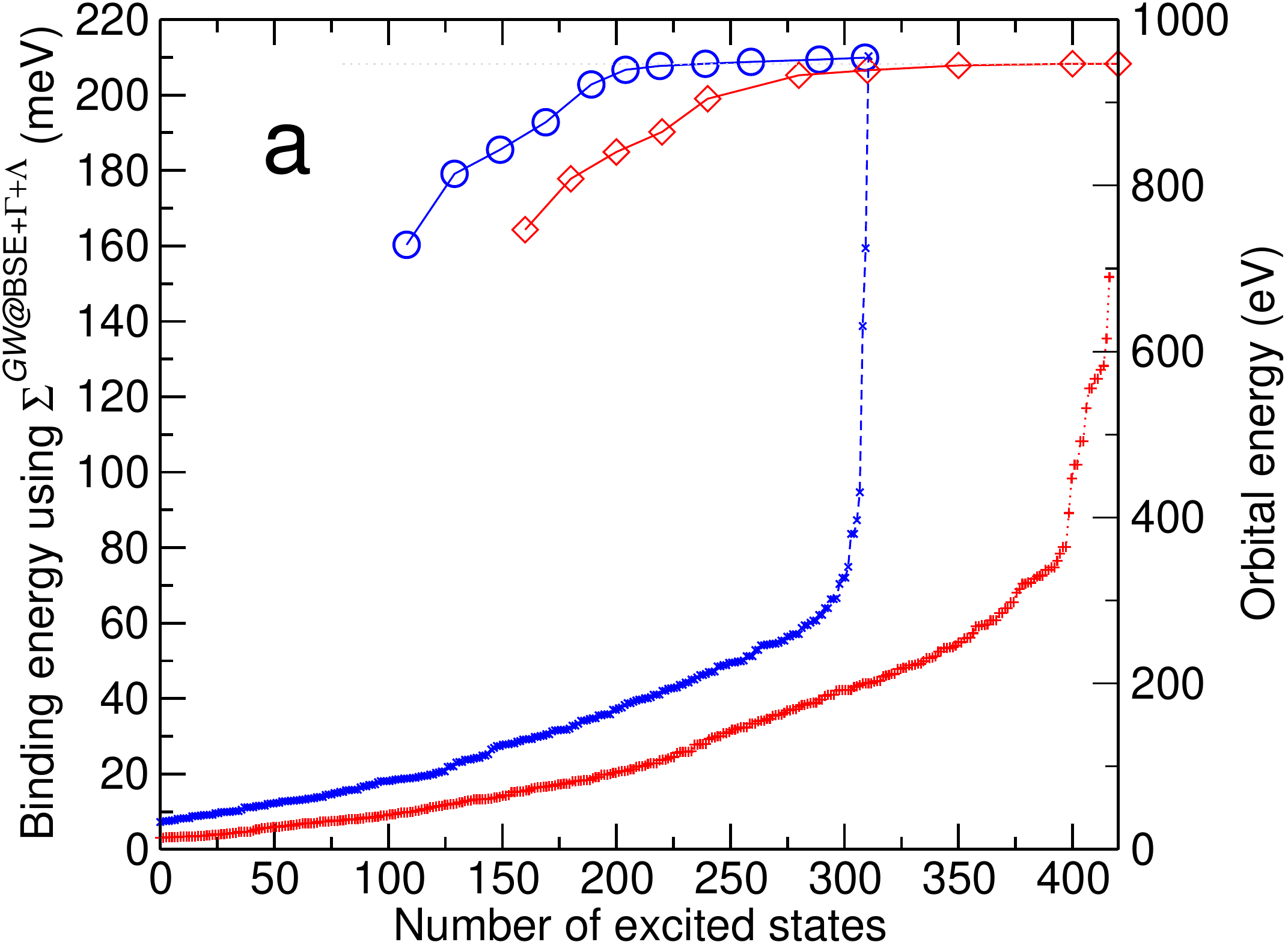}~~~
\includegraphics*[width=0.49\textwidth]{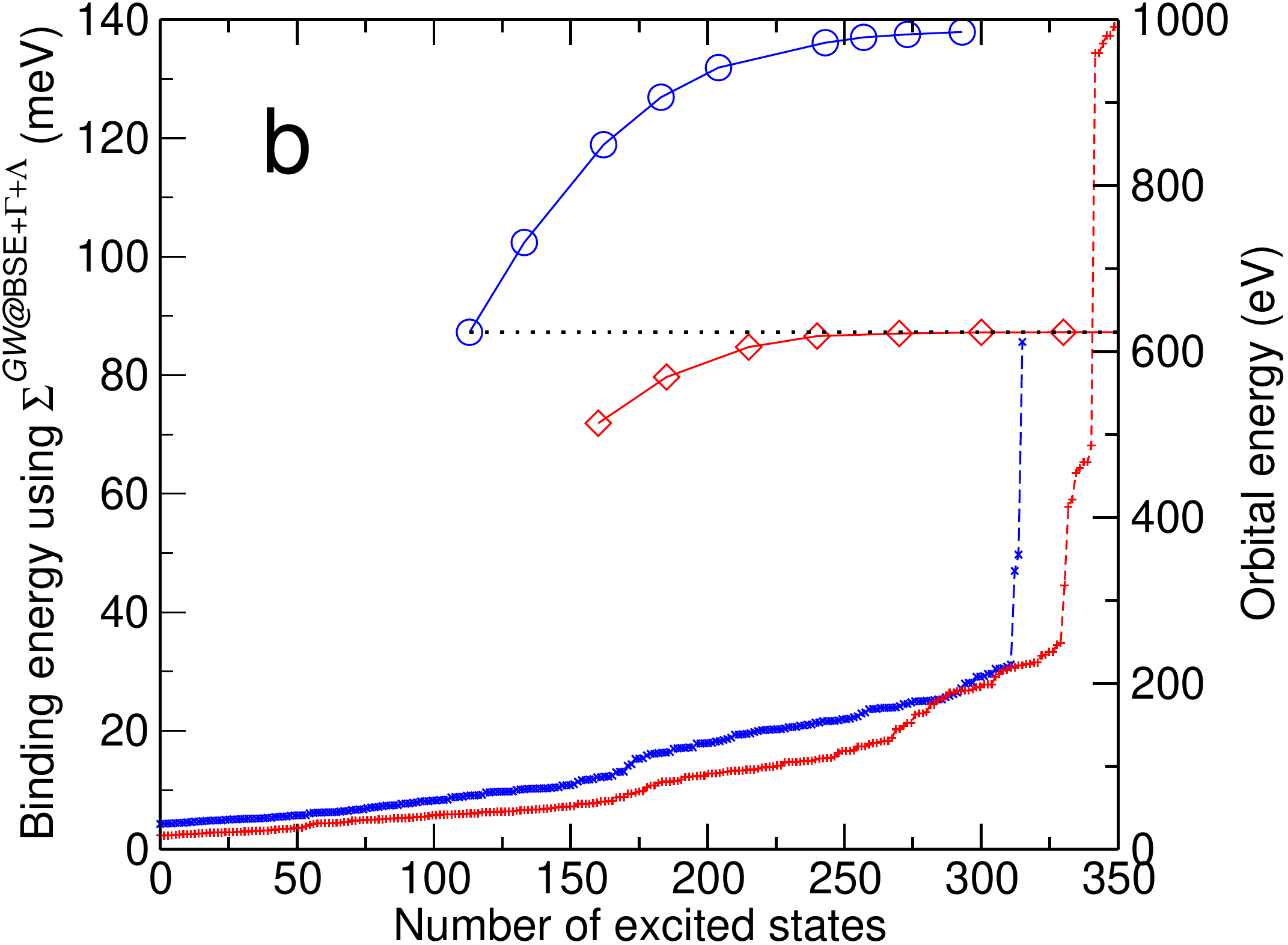}
\caption*{\footnotesize{\bf Extended Data Figure 2. Convergence of positron binding energies in acetonitrile and CSe$_2$ with respect to electron and positron basis size}.
\footnotesize{Positron binding energy calculated using the $\Sigma^{GW@{\rm BSE}+\Gamma+\Lambda}$ self energy for varying number of electron (positron) HF MOs (whose energies are shown as blue and red crosses, respectively) included in the basis. % for the many-body calculations. 
For acetonitrile, the varying electron (positron) MO calculations included all positron (electron) MOs. For CSe$_2$, the varying electron MO calculations included all positron MOs, whilst the varying positron MO calculation included 113 electron MOs (indicated by the lowest blue circle). 
The binding energy reaches convergence when the electronic orbital with energies up to $\sim$150--200 eV are included. Similar behaviour was also observed for the other molecules considered.
%, though for acetone and propanal convergence is slower. 
%In these calculations, all of the positron excited states (whose energies shown as red symbols) were included.
}
\label{fig:acetocon}}
\end{figure}

\renewcommand{\figurename}{\small {\bf Extended Data Figure}}
\begin{figure}[hp!!]
\centering
\includegraphics*[width=0.5\textwidth]{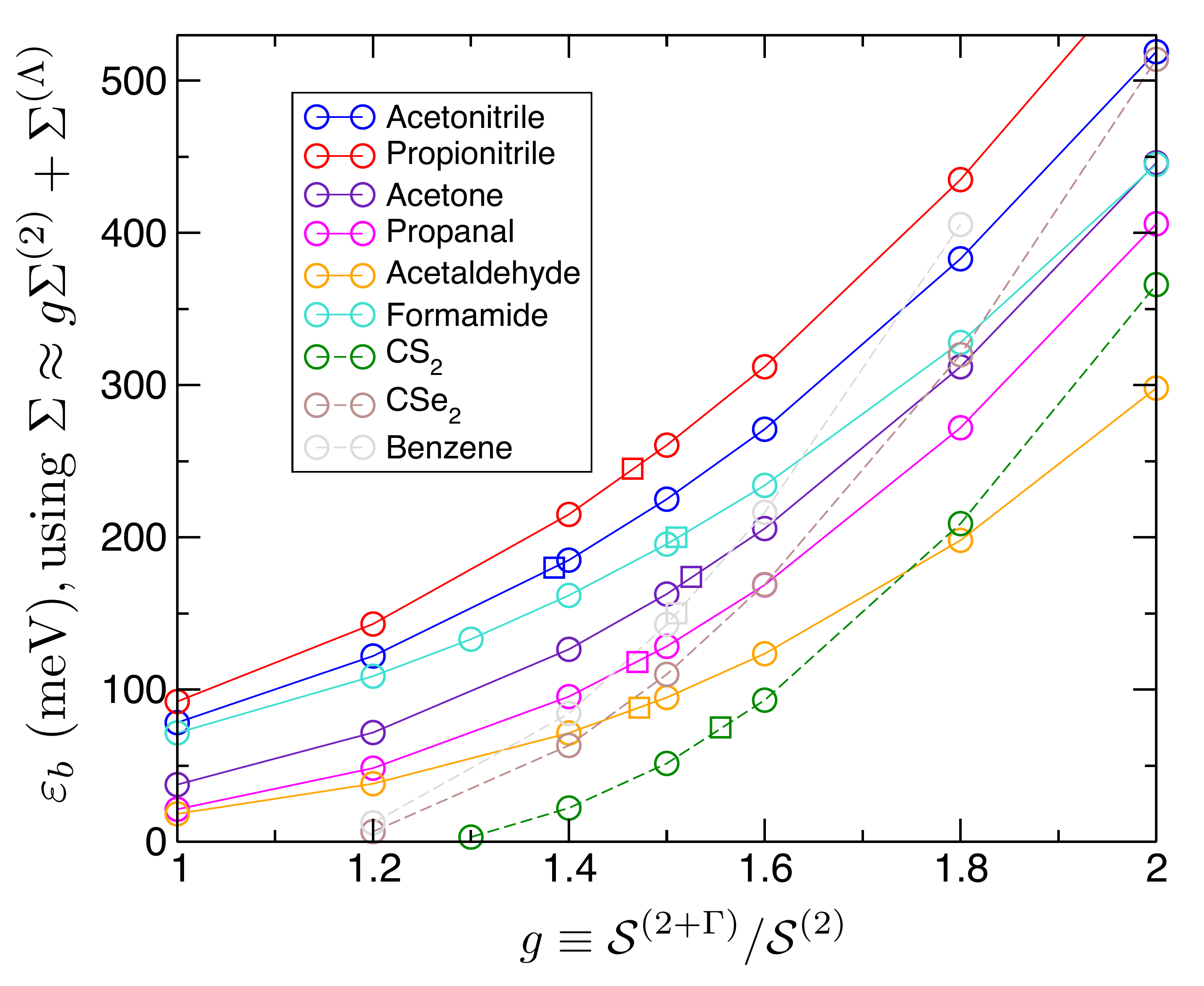}~~~~
\caption*{\footnotesize{\bf Extended Data Figure 3. Non-linearity of the binding energy and strength of correlation potential.}
\footnotesize{Binding energy calculated approximating the positron self energy $\Sigma$ as $\Sigma\approx g\Sigma^{(2)}+\Sigma^{\Lambda}$ as a function of the scaling parameter $g\equiv \mathcal{S}^{2+\Gamma}/\mathcal{S}^{(2)}$ (see main text for more details) (circles). Experiment (squares) is from [\citen{Danielson10,Danielson12}]; for formamide preliminary measurements find a binding energy of $\varepsilon_b\sim200$\,meV, but a final result is yet to be determined. See also Fig.~3\,c of main text.  
}
\label{fig:acetocon}}
\end{figure}

\renewcommand{\figurename}{\small {\bf Extended Data Figure}}
\begin{figure}[hp!!]
\centering
\includegraphics*[width=0.98\textwidth]{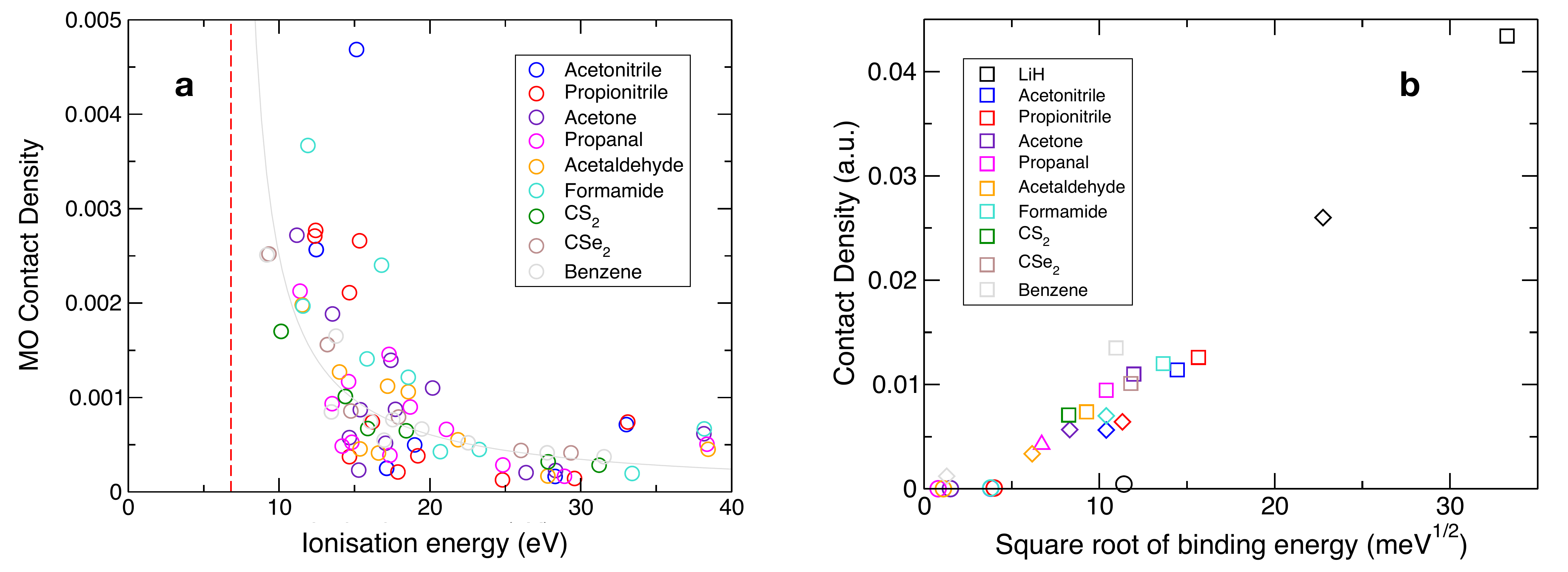}
\caption*{\footnotesize{{\bf Extended Data Figure 4. Calculated electron-positron contact density}. {\bf a}, contact density  for individual electronic MOs as a function of their ionisation energy, calculated  including vertex enhancement factors and renormalization coefficients (see `Methods' Eqns.~6 and 7). Red dashed line: positronium ground state energy at $|E_{\rm Ps}|=6.8$ eV. Grey line: $\delta_{ep}=0.008/(I-|E_{\rm Ps}|)$ (for a guide). Also see Extended Data Table 3 and Fig.~4 of main text. 
{\bf b}, Contact density calculated at the HF (circles) and  various levels of many-body theory (diamonds: $GW$@BSE; squares: $GW$@BSE+$\Gamma$+$\Lambda$) against the square root of the binding energy.}\\[4ex]
%\includegraphics[width=1\textwidth]{lifetimes} 
%\caption*{\footnotesize {\bf Extended Data Figure 5. Calculated electron-positron contact densities and positron lifetimes with respect to annihilation.}  
%{\bf a}, fractional contribution of individual MOs to the total electron-positron contact density (Eqn.~6 in `Methods).
%%with x-axis counting down from the HOMO (highest-occupied molecular orbital) 
%{\bf b} and {\bf c}, the electron-positron contact density (magenta) at the $\Sigma^{GW+\Gamma+\Lambda}$ level for the HOMO and (H-1)OMO
% in acetonitrile (blue and brown show positive and negative electron wavefunction regions, respectively), c.f., Fig.~2 (b) (positron density).
%{\bf Table: Positron lifetimes with respect to annihilation}. $\tau^{(0)}$: lifetime calculated in the Hartree-Fock independent particle approximation excluding the vertex enhancement factors and using a positron wavefunction normalised to unity; $\tau^{GW}$ and $\tau$: lifetime calculated using the Dyson positron wavefunction at the $\Sigma^{GW}$ and $\Sigma^{GW+\Gamma+\Lambda}$ levels including vertex enhancement factors and renormalization constants (Eqns.~6 and 7 of `Methods'). 
%}\label{fig:cds}
\label{fig:acetocon}}
\end{figure}

\newpage
\footnotesize
\bibliographystyle{naturemag}
%\bibliography{posmol-mbt-binding.bib}

\end{document}